\documentclass[apj]{emulateapj}

\usepackage{apjfonts,ulem}
\usepackage{amssymb, amsmath}
\newcommand\degree{\degr}
\newcommand\degrees\degree

\DeclareSymbolFont{UPM}{U}{eur}{m}{n}
\DeclareMathSymbol{\umu}{0}{UPM}{"16}
\let\oldumu=\umu
\renewcommand\umu{\ifmmode\oldumu\else\math{\oldumu}\fi}
\newcommand\micro{\umu}
\renewcommand\micron{\micro m}
\newcommand\microns \micron

\let\oldsim=\sim
\renewcommand\sim{\ifmmode\oldsim\else\math{\oldsim}\fi}
\let\oldpm=\pm
\renewcommand\pm{\ifmmode\oldpm\else\math{\oldpm}\fi}
\newcommand\by{\ifmmode\times\else\math{\times}\fi}
\newcommand\ttt[1]{10\sp{#1}}
\newcommand\tttt[1]{\by\ttt{#1}}

\newbox{\wdbox}
\renewcommand\c{\setbox\wdbox=\hbox{,}\hspace{\wd\wdbox}}
\renewcommand\i{\setbox\wdbox=\hbox{i}\hspace{\wd\wdbox}}

\newcount\timect
\newcount\hourct
\newcount\minct
\newcommand\now{\timect=\time \divide\timect by 60
         \hourct=\timect \multiply\hourct by 60
         \minct=\time \advance\minct by -\hourct
         \number\timect:\ifnum \minct < 10 0\fi\number\minct}

\catcode`@=11

\newcommand\comment[1]{}
\newcommand\commenton{\catcode`\%=14}
\newcommand\commentoff{\catcode`\%=12}

\renewcommand\math[1]{$#1$}
\newcommand\mathshifton{\catcode`\$=3}
\newcommand\mathshiftoff{\catcode`\$=12}

\comment{the backslash is necessary}

\comment{alignment tab}

\let\atab=&
\newcommand\atabon{\catcode`\&=4}
\newcommand\ataboff{\catcode`\&=12}

\let\oldmsp=\sp
\let\oldmsb=\sb
\def\sp#1{\ifmmode
           \oldmsp{#1}%
         \else\strut\raise.85ex\hbox{\scriptsize #1}\fi}
\def\sb#1{\ifmmode
           \oldmsb{#1}%
         \else\strut\raise-.54ex\hbox{\scriptsize #1}\fi}
\newbox\@sp
\newbox\@sb
\def\sbp#1#2{\ifmmode%
           \oldmsb{#1}\oldmsp{#2}%
         \else
           \setbox\@sb=\hbox{\sb{#1}}%
           \setbox\@sp=\hbox{\sp{#2}}%
           \rlap{\copy\@sb}\copy\@sp
           \ifdim \wd\@sb >\wd\@sp
             \hskip -\wd\@sp \hskip \wd\@sb
           \fi
        \fi}
\def\msp#1{\ifmmode
           \oldmsp{#1}
         \else \math{\oldmsp{#1}}\fi}
\def\msb#1{\ifmmode
           \oldmsb{#1}
         \else \math{\oldmsb{#1}}\fi}
\def\supon{\catcode`\^=7}
\def\supoff{\catcode`\^=12}
\def\subon{\catcode`\_=8}
\def\suboff{\catcode`\_=12}
\def\supsubon{\supon \subon}
\def\supsuboff{\supoff \suboff}

\newcommand\actcharon{\catcode`\~=13}
\newcommand\actcharoff{\catcode`\~=12}

\newcommand\paramon{\catcode`\#=6}
\newcommand\paramoff{\catcode`\#=12}

\comment{And now to turn us totally on and off...}

\newcommand\reservedcharson{\commenton \mathshifton \atabon \supsubon \actcharon
	\paramon}

\newcommand\reservedcharsoff{\commentoff \mathshiftoff \ataboff
	\supsuboff \actcharoff \paramoff}

\catcode`@=12
\reservedcharsoff

\reservedcharson

\comment{ Must have ONLY ONE of these... trust these macros, they work

}

\newcommand{\squishlist}{
 \begin{list}{$\bullet$}
  { \setlength{\itemsep}{1pt}
     \setlength{\parsep}{0pt}
     \setlength{\topsep}{3pt}
     \setlength{\partopsep}{0pt}
     \setlength{\leftmargin}{2.0em}
     \setlength{\labelwidth}{1.5em}
     \setlength{\labelsep}{0.5em} } }

\newcommand{\squishend}{
  \end{list}  }

\reservedcharson
\actcharon

\usepackage[colorlinks=true,citecolor=blue,urlcolor=cyan]{hyperref}
\usepackage{rotating}
\shorttitle{{\em Spitzer} Phase Curve Constraints for WASP-43b at 3.6 and 4.5~{\microns}}
\shortauthors{Stevenson {\em et al.}}

\submitted{}

\begin{document}

\title{{\em Spitzer} Phase Curve Constraints for WASP-43\lowercase{b} at 3.6 and 4.5~{\microns}}

\author{Kevin B.\ Stevenson\altaffilmark{1,2}}
\author{Michael R.\ Line\altaffilmark{3,4}}
\author{Jacob L.\ Bean\altaffilmark{1}}
\author{Jean-Michel D\'esert\altaffilmark{5}}
\author{Jonathan J.\ Fortney\altaffilmark{6}}
\author{Adam P.\ Showman\altaffilmark{7}}
\author{Tiffany Kataria\altaffilmark{8,9}}
\author{Laura Kreidberg\altaffilmark{1,10,11}}
\author{Y. Katherina Feng\altaffilmark{6}}
\affil{\sp{1}Department of Astronomy and Astrophysics, University of Chicago, Chicago, IL 60637, USA}
\affil{\sp{2}Space Telescope Science Institute, Baltimore, MD 21218, USA}
\affil{\sp{3}NASA Ames Research Center, Moffet Field, CA 94035, USA}
\affil{\sp{4}School of Earth \& Space Exploration, Arizona State University, Tempe, AZ 85287, USA}
\affil{\sp{5}Anton Pannekoek Institute for Astronomy, University of Amsterdam, The Netherlands}
\affil{\sp{6}Department of Astronomy and Astrophysics, University of California, Santa Cruz, CA 95064, USA}
\affil{\sp{7}Department of Planetary Sciences and Lunar and Planetary Laboratory, University of Arizona, Tucson, AZ 85721, USA}
\affil{\sp{8}Astrophysics Group, School of Physics, University of Exeter, Stocker Road, Exeter EX4 4QL, UK}
\affil{\sp{9}NASA Jet Propulsion Laboratory, Pasadena, CA 91109, USA}
\affil{\sp{10}Harvard-Smithsonian Center for Astrophysics, Cambridge, MA 02138, USA}
\affil{\sp{11}Harvard Society of Fellows, Harvard University, Cambridge, MA 02138, USA}

\email{E-mail: kbs@stsci.edu}

\begin{abstract}
Previous measurements of heat redistribution efficiency (the ability to transport energy from a planet's highly irradiated dayside to its eternally dark nightside) show considerable variation between exoplanets.  Theoretical models predict a positive correlation between heat redistribution efficiency and temperature for tidally locked planets; however, recent {\em Hubble Space Telescope (HST)} WASP-43b spectroscopic phase curve results are inconsistent with current predictions.  Using the {\em Spitzer Space Telescope}, we obtained a total of three phase curve observations of WASP-43b ($P = 0.813$~days) at 3.6 and 4.5~{\microns}.  The first 3.6~{\micron} visit exhibits spurious nightside emission that requires invoking unphysical conditions in our cloud-free atmospheric retrievals.  The two other visits exhibit strong day-night contrasts that are consistent with the {\em HST} data.  To reconcile the departure from theoretical predictions, WASP-43b would need to have a high-altitude, nightside cloud/haze layer blocking its thermal emission.  Clouds/hazes could be produced within the planet's cool, nearly retrograde mid-latitude flows before dispersing across its nightside at high altitudes.  Since mid-latitude flows only materialize in fast-rotating ($\lesssim1$ day) planets, this may explain an observed trend connecting measured day-night contrast with planet rotation rate that matches all current {\em Spitzer} phase curve results.  Combining independent planetary emission measurements from multiple phases, we obtain a precise dayside hemisphere H\sb{2}O abundance ($2.5\tttt{-5} - 1.1\tttt{-4}$ at 1$\sigma$ confidence) and, assuming chemical equilibrium and a scaled solar abundance pattern, we derive a corresponding metallicity estimate that is consistent with being solar (0.4 -- 1.7).  Using the retrieved global CO+CO\sb{2} abundance under the same assumptions, we estimate a comparable metallicity of 0.3 -- 1.7$\times$ solar.  This is the first time that precise abundance and metallicity constraints have been determined from multiple molecular tracers for a transiting exoplanet.
\end{abstract}

\section{Introduction}
\label{intro}

Exoplanet phase curves provide a wealth of information about planetary atmospheres, including longitudinal constraints on atmospheric composition, thermal structure, and energy transport.  In the thermal infrared, the amplitude of the phase variation determines the day-night temperature contrast and the offset determines the longitude of the planet's hottest point.  The amplitude and offset derive primarily from equatorial jets redistributing heat from the hot dayside to the cooler nightside and are non-trivially connected to an atmosphere's radiative, advective/vertical, and drag timescales \citep{Cowan2011, Perna2012, Perez-Becker2013, Crossfield2015b, Komacek2016}.  Constraints on these properties allow us to begin understanding the fundamental processes occurring in highly irradiated atmospheres.

The presence of clouds/hazes in exoplanet atmospheres is not well understood \citep[e.g., ][]{Morley2015, Sing2016, Stevenson2016b} and their effect on phase curve observations is unknown.  If present on the planet dayside, condensates in a hot-Jupiter atmosphere can modify the measured redistribution of heat \citep{Pont2013}.  To first order, the presence of clouds/hazes moves the infrared photosphere to higher altitudes (lower pressures) where the radiative timescales are shorter.  This, in turn, increases the measured day-night contrast and reduces the phase curve peak offset \citep{Sudarsky2003}.  However, cloud inhomogeneities (or patchiness) can weaken this effect \citep{Parmentier2016}.  If present on the planet nightside, obscuring clouds/hazes similarly modify the observable photosphere to higher altitudes, again increasing the measured day-night contrast \citep{Kataria2015}.  A more in-depth understanding of the effects of clouds requires high-precision spectrophotometric observations at all orbital phases.

\subsection{Previous Results}

In 2011, \citet{Hellier2011} announced the detection of a hot-Jupiter exoplanet orbiting a K7 star, WASP-43.  The relative sizes and temperatures of these two bodies results in relatively deep eclipse depths that are favorable for exoplanet characterization.  When combined with WASP-43b's short, 19.5~hour orbital period, this has encouraged multiple, ground-based observational campaigns in both the optical and near infrared.  

\citet{Gillon2012} obtained nearly two dozen ground-based transit light curves to improve the precision of many system parameters.  They also reported a high-confidence detection of thermal emission at 2.09~{\microns} (1560 {\pm} 140 ppm).  \citet{Wang2013} observed WASP-43b during secondary eclipse and published $H$ and $K_s$-band detections of 1030 {\pm} 170 ppm and 1940 {\pm} 290 ppm, respectively.  Similarly, \citet{Chen2014} reported a $K$-band detection of 1970 {\pm} 420 ppm.  Although generally consistent with each other, all of these ground-based detections of thermal emission are inconsistent with the high-precision {\em HST}/WFC3 eclipse depths and best-fit atmospheric models presented by \citet{Stevenson2014c}.  This strengthens recent concern that ground-based observations may tend to over-predict measured eclipse depths in lower-quality light curves due to unexplained or under-modeled systematics \citep{Rogers2013}.

Using the {\em Spitzer Space Telescope}, \citet{Blecic2014} measured the dayside emission of WASP-43b at 3.6 and 4.5~{\microns}.  Consistent with previous studies, they ruled out the presence of a strong thermal inversion, suggested low day-night heat redistribution, and found that atmospheric models assuming an oxygen-rich composition achieve the best fits to the available data.  Due to WASP-43b's proximity to its host star, \citet{Blecic2014} also attempted to estimate the decay rate of the planet's orbital period.  They determined that the measured period change ($\dot{p} = -0.095 {\pm} 0.036$~s~yr\sp{-1}) was not significant.

\citet{Murgas2014} obtained long-slit spectra of WASP-43 in the red optical over five full- or partial-transit observations.  Their measured transmission spectrum contains a weak excess near the Na I doublet and a smoothly varying trend at redder wavelengths.  \citet{Murgas2014} also placed constraints on tidal decay rate of WASP-43b, finding a value ($\dot{p} = -0.15 {\pm} 0.06$~s~yr\sp{-1}) that is consistent with that reported by \citet{Blecic2014}.  However, with seven additional transit timing constraints, \citet{Ricci2015} found no evidence of orbital decay.

In \citet{Stevenson2014c}, we present spectroscopic thermal emission measurements of WASP-43b as a function of orbital phase.  {\em HST}/WFC3 acquired data that spanned three full planet rotations, plus three primary transits and two secondary eclipses.  Our analyses confirm previous reports of low day-night heat redistribution and contrast with the modest day-night differences inferred from {\em Spitzer} photometric phase curves of similarly irradiated giant planets \citep{Perez-Becker2013}.  The band-integrated phase curve exhibits a strong asymmetry where the emission maximum occurs 40 {\pm} 3 minutes prior to the midpoint of secondary eclipse and the minimum occurs 34 {\pm} 5 minutes after the primary transit midpoint.  Best-fit atmospheric models favor the presence of H\sb{2}O and a monotonically decreasing temperature with pressure at all longitudes.  We also uncovered an altitude dependence in the hotspot offset relative to the substellar point that is qualitatively consistent with brown dwarf measurements and circulation-model predictions.

In the first of two companion papers, we constrained the abundance of water using both the transmission and dayside emission spectra \citep{Kreidberg2014b}.  The derived water content is consistent with solar composition (0.4 -- 3.5$\times$ solar) and the inferred metallicity matches the trend observed in the solar system giant planets wherein more massive bodies have lower metal enrichment.

In the second companion paper, we presented 3D atmospheric circulation models of WASP-43b that explored the effects of composition, metallicity, and frictional drag \citep{Kataria2015}.  We found that a 5$\times$  solar metallicity model provides a good match to the dayside emission spectrum and exhibits equatorial superrotation that explains the observed eastward-shifted hotspot.  We noted, however, that the model nightside is brighter than that observed with {\em HST}/WFC3 and suggested that the existence of thick, high-altitude clouds on the planet nightside could lower the measured flux and resolve the discrepancy.

\subsection{Roadmap}

In this paper, we present full-orbit, photometric phase curves of WASP-43b obtained at 3.6 and 4.5~{\microns} by the {\em Spitzer Space Telescope}.  Because the planet is presumed to be tidally locked, where its rotation rate is equal to its orbital period, these measurements constrain WASP-43b's emission as a function of planet longitude.  This, in turn, provides insight into how efficiently the planet's atmosphere transports heat from its irradiated dayside to its permanent nightside.

In Section \ref{sec:obs}, we discuss the acquisition and reduction of {\em Spitzer} data, how we handle the position- and time-dependent systematics in our light-curve model fits, and uncertainty estimation.  Section \ref{sec:results} presents the results of our {\em Spitzer} analysis and compares them to previous work.  In Section \ref{sec:atm}, we combine our results with those from our {\em HST}/WFC3 analysis \citep{Stevenson2014c} to place tighter constraints on the planet's atmospheric composition, metallicity, and thermal structure.  Section \ref{sec:circ} compares the measured heat redistribution efficiency to that of other planets and considers correlations with equilibrium temperature and planet rotation rate.  Finally, we summarize our findings in Section \ref{sec:concl}.

\section{OBSERVATIONS AND DATA ANALYSIS}
\label{sec:obs}

\subsection{Observations and Reduction}

{\em Spitzer} obtained three broadband photometric phase curves of WASP-43b (Programs 10169 \& 11001, PI: Kevin Stevenson), each lasting 25.4 hours.  Using the subarry mode with two-second frame times, the InfraRed Array Camera \citep[IRAC,][]{IRAC} acquired two datasets at 3.6~{\microns} and a single dataset at 4.5~{\microns}.  Table \ref{tab:ObsDates} provides specific details for each observation.

\begin{figure}[tb]
\centering
\includegraphics[width=1.0\linewidth,clip]{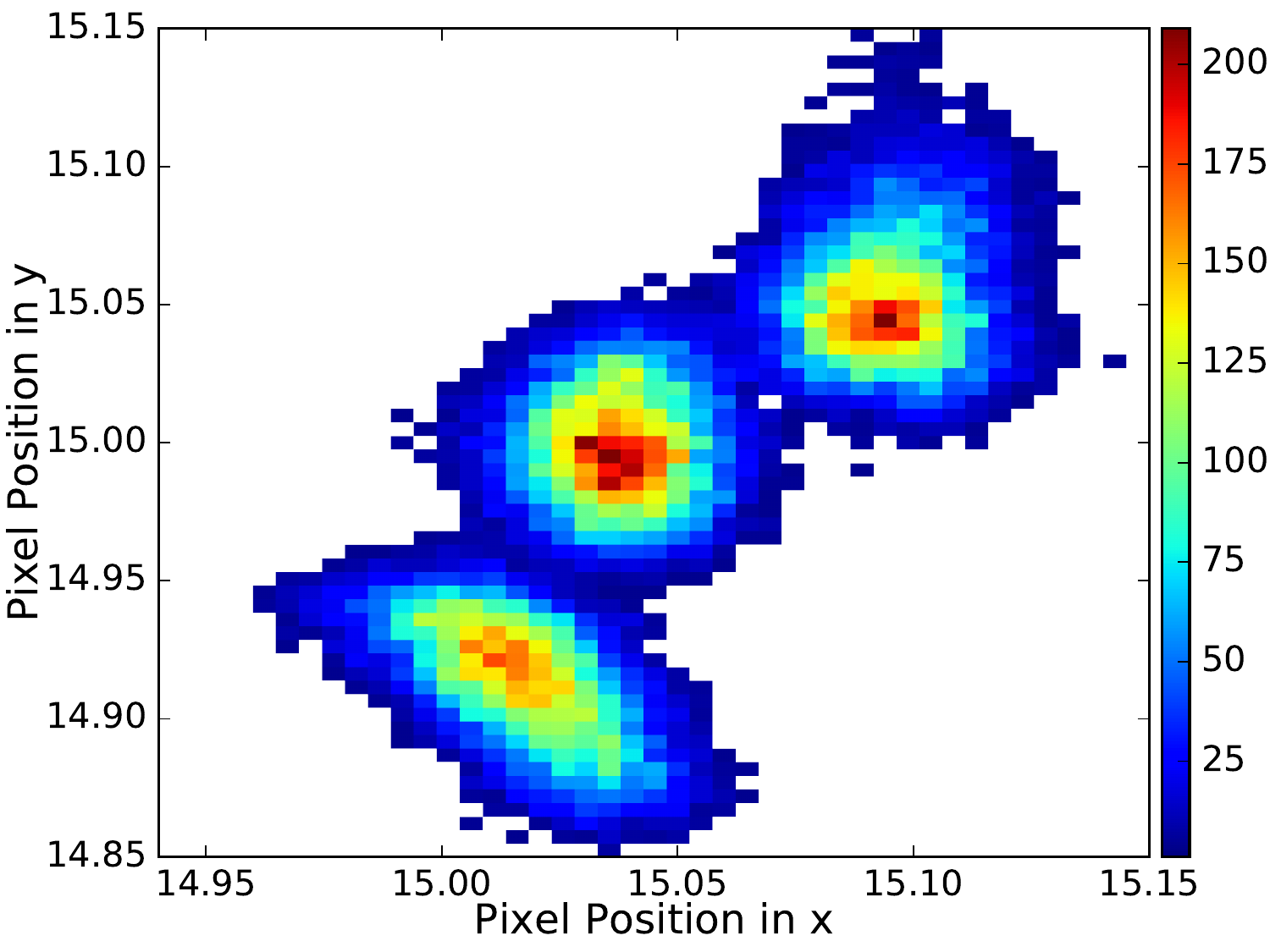}
\includegraphics[width=1.0\linewidth,clip]{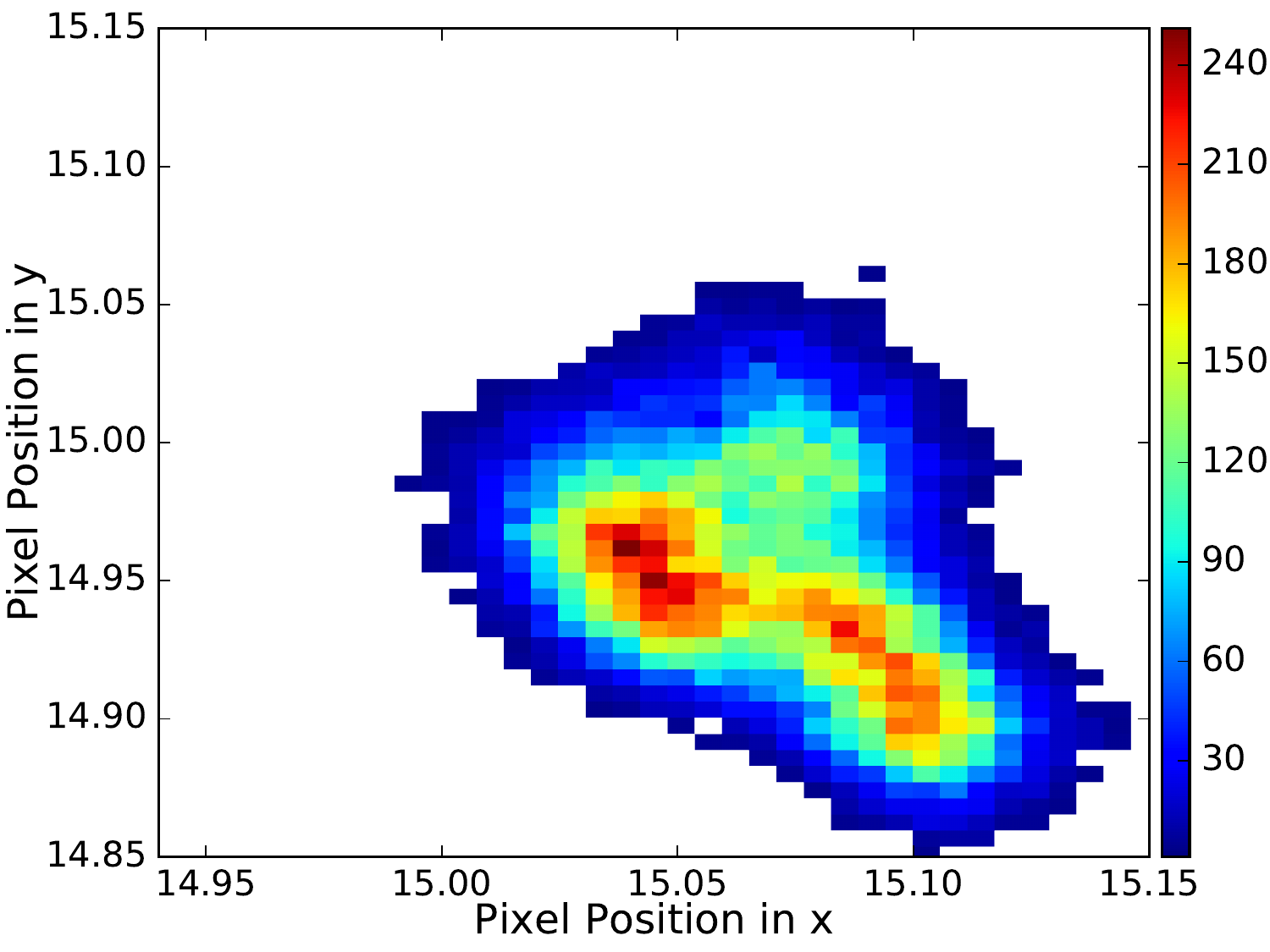}
\caption{\label{fig:pointingHist}{
Pointing histograms for the first (wa043bo11, top) and second (wa043bo12, bottom) 3.6~{\micron} visits.  The three distinct regions in the top panel are due to repointing inconsistencies at the start of each AOR.  Nominally, pointing corrections should return the telescope to overlapping positions (see bottom panel), thus limiting the effects of telescope drift and enhancing our ability to model the position-dependent systematics.
}}
\end{figure}

Before initiating our science program, we employed a standard 30-minute pre-observation using the PCRS peak-up to mitigate spacecraft drift.  Science observations commenced at a planetary orbital phase of $\sim$0.35 and were initially divided into three, 8.5 hour astronomical observation requests (AORs) to minimize long-term spacecraft drift by repositioning our target onto {\em Spitzer's} defined sweet spot in subarray mode.  This strategy worked well at 4.5~{\microns}; however, repointing during the first 3.6~{\micron} visit produced inconsistent results and the measured centroids from each AOR exhibit minimal overlap (see Figure \ref{fig:pointingHist}).  As discussed in Section \ref{sec:3.6results}, the first 3.6~{\micron} visit exhibited strong nightside planetary emission that required invoking unphysical conditions in our cloud-free atmospheric retrievals.  Since the second AOR contains information on the planet's nightside emission but has no secondary eclipse to act as an anchor, we suspected the minimal centroid overlap between AORs to be the source of the discrepant nightside flux.  Therefore, we requested (and were granted) time for a second 3.6~{\micron} phase curve observation.  This time, we changed our strategy by dividing the visit into two AORs, the first being 15.2 hours in duration and the second being 10.2 hours.  The reduction in the number of AORs improved phase-curve accuracy by providing anchor points during secondary eclipse for each AOR.  The AOR durations were asymmetric to avoid starting a new AOR during primary transit.

To reduce the data, we used the Photometry for Orbits, Eclipses, and Transits (POET) pipeline \citep{Campo2011,Stevenson2011,Cubillos2013}.  A general description of the reduction process is as follows.  POET flags bad pixels using a double-iteration, 4$\sigma$ filter at each pixel column in our stack of 64 subarray frames, determines image centers from a 2D Gaussian fit \citep{Lust2014}, and applies $5\times$ interpolated aperture photometry \citep{Harrington2007} over a range of aperture sizes in 0.25 pixel increments.  In addition to our phase curves, we also reanalyzed WASP-43b eclipse observations from {\em Spitzer} Program 70084 (PI: Joseph Harrington).

\begin{table*}[tb]
\centering
\caption{\label{tab:ObsDates} 
Observation Information}
\begin{tabular}{crccclcc}
    \hline
    \hline
    Label\tablenotemark{a}       
                & Observation Date      & Duration  & Frame Time    & Total Frames  & {\em Spitzer} & Wavelength    & Previous      \\
                &                       & (hours)   & (seconds)     &               & Pipeline      & (\microns)    & Publication   \\
    \hline
    wa043bs21   &         2011 July 29  & 5.9       & 2.0           & 10496         & S18.18.0      & 4.5           & \citet{Blecic2014}\\
    wa043bs11   &         2011 July 30  & 5.9       & 2.0           & 10496         & S18.18.0      & 3.6           & \citet{Blecic2014}\\
    wa043bo21   & 2014 August 27 -- 28  & 25.4      & 2.0           & 44928         & S19.1.0       & 4.5           & ---      \\
    wa043bo11   &    2015 March 7 -- 8  & 25.4      & 2.0           & 44928         & S19.1.0       & 3.6           & ---      \\
    wa043bo12   & 2015 September 4 -- 5 & 25.4      & 2.0           & 44928         & S19.2.0       & 3.6           & ---      \\
    \hline
\end{tabular}
\tablenotetext{1}{wa043b designates the planet, $s$/$o$ specifies secondary eclipse or orbital phase curve, and \#\# identifies the wavelength and observation number.}
\end{table*}

\subsection{Light-Curve Systematics and Fits}

After a decade of work, {\em Spitzer's} systematics are generally considered to be well understood \citep[e.g.][]{Charbonneau2005, Agol2010, Knutson2011, Ingalls2012, Stevenson2011, Lewis2013, Deming2015}.  Most groups now agree on a common set of best practices and, in a recent IRAC Data Challenge, most pipelines produced consistent results when testing against real and artificial datasets \citep{Ingalls2016}.  Therefore, although older analyses may warrant some degree of skepticism \citep{Hansen2014}, newer results are becoming more reliable.

The dominant systematic at 3.6 and 4.5~{\microns} is a position-dependent flux that is sensitive to intra-pixel (IP) variations at the hundredth-of-a-pixel scale.  We apply Bilinearly-Interpolated Subpixel Sensitivity (BLISS) mapping \citep{Stevenson2011} to model position-dependent systematics, but also tested Pixel-Level Decorrelation \citep[PLD,][]{Deming2015} with the phase curve observations.  We find that the PLD method with linear coefficients does not adequately correct the intra-pixel effect in some regions of pixel space, thus achieving slightly worse fits overall.  The poor fit may be due to the relatively large variations in pixel position over the duration of the phase curve observations.  {\em Spitzer} light curves sometimes exhibit a weak, visit-long trend that we model using a linear or quadratic function in time.  We reject model combinations (including those with no visit-long ramp) that have higher Bayesian Information Criterion \citep[BIC,][]{Liddle2008} values (see Table \ref{tab:Models}).

\begin{table}[tb]
\centering
\caption{\label{tab:Models} 
Reduction and Light Curve Model Components}
\begin{tabular}{ccccc}
    \hline
    \hline
    Label       & Aperture Size & IP Model      & Ramp Model    & $\Delta$BIC   \\
                & (pixels)      &               &               &               \\
    \hline
    wa043bs21   & 2.5           & BLISS         & --            & 38.2          \\
                & "             & "             & Linear        &  0.0          \\
                & "             & "             & Quadratic     &  9.2          \\
    wa043bs11   & 2.5           & BLISS         & --            & 50.9          \\
                & "             & "             & Linear        & 16.7          \\
                & "             & "             & Quadratic     &  0.0          \\
    wa043bo21   & 2.0           & BLISS         & --            & 54.5          \\
                & "             & "             & Linear        &  0.0          \\
                & "             & "             & Quadratic     &  3.4          \\
    wa043bo11   & 2.5           & BLISS         & --            &  0.0          \\
                & "             & "             & Linear        &  5.6          \\
                & "             & "             & Quadratic     &  7.6          \\
    wa043bo12   & 3.0           & BLISS         & --            & 111.6         \\
                & "             & "             & Linear        &   0.0         \\
                & "             & "             & Quadratic     &   8.1         \\
    \hline
\end{tabular}
\end{table}

To model the planet's emission as a function of orbital phase, we use a sinusoidal function of the form
\begin{equation}
c_1\cos[2\pi(t-c_2)/P]+c_3\cos[4\pi(t-c_4)/P],
\end{equation}
\noindent where $t$ is time, $P$ is the planet's orbital period, and $c_1$ - $c_4$ are free parameters.  The second sinusoidal term allows us to fit for an asymmetric phase curve.  Similar formulations have been used to model other exoplanet phase curves \citep[e.g.][]{Cowan2008, Knutson2012, Lewis2013, Stevenson2014c}.

To fit the shapes of primary transit and secondary eclipse, we follow the prescription of \citet{Mandel2002}.  The former requires the application of a stellar limb-darkening model, for which we adopt a quadratic equation \citep{Claret2000} with values derived from stellar Kurucz models \citep{Kurucz2004}.  For WASP-43, these values are (0.10910, 0.17577) at 3.6~{\microns} and (0.10092, 0.12797) at 4.5~{\microns}.
  
We find the best solution by fitting all of the free parameters simultaneously using a Levenberg-Marquardt minimizer.  We estimate parameter uncertainties using a Differential-Evolution Markov Chain algorithm \citep[DEMC,][]{terBraak2008}.  All of the 3.6~{\micron} datasets exhibit a fair amount of time-correlated noise; therefore, we include these effects in our uncertainty estimates using the wavelet analysis described by \citet{Carter2009b}.  Neither 4.5~{\micron} dataset requires this additional step.

\section{RESULTS}
\label{sec:results}

\begin{figure*}[t]
\centering
\includegraphics[width=1.0\linewidth]{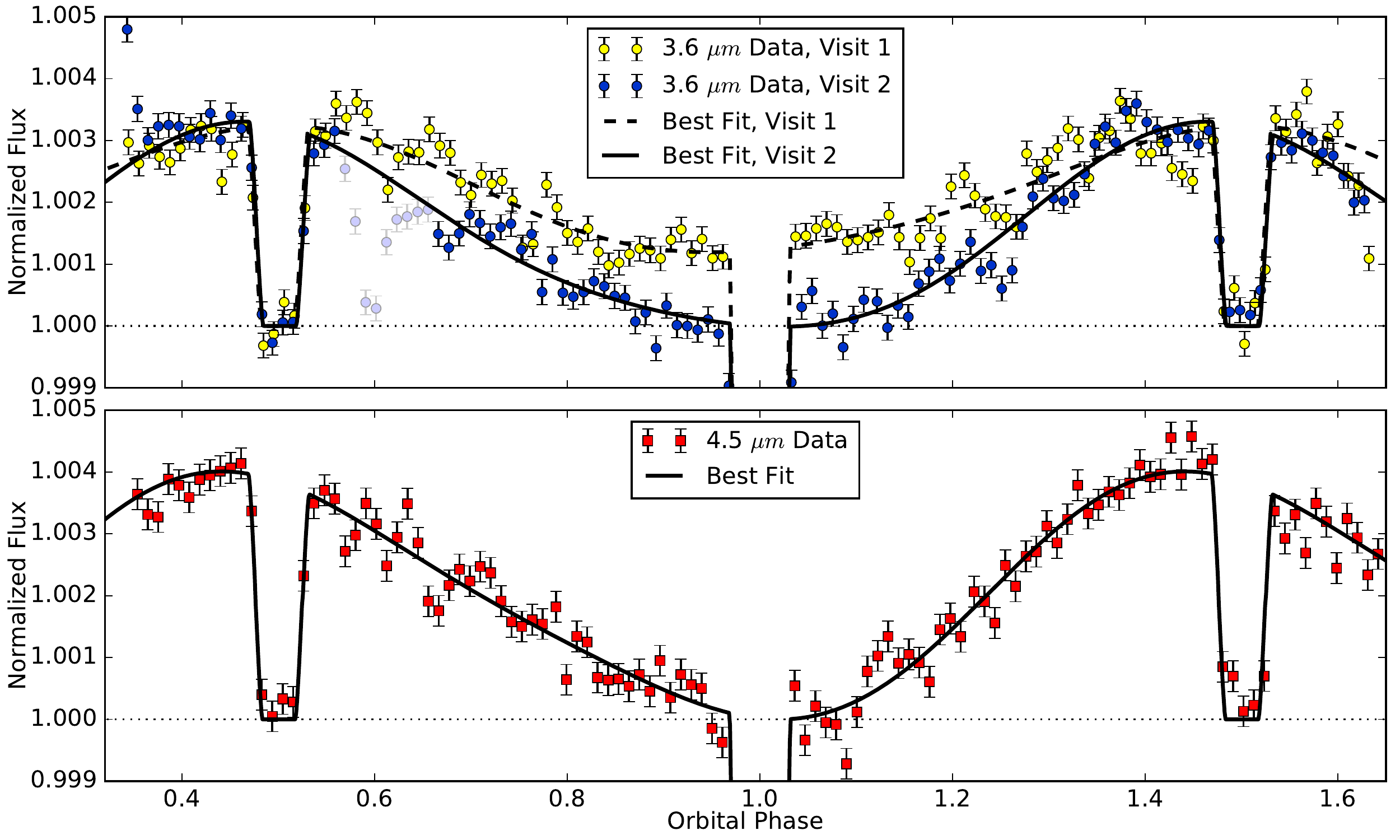}
\caption{\label{fig:PhaseCurves}{
Emission phase curves of WASP-43b at 3.6 (top) and 4.5 (bottom)~{\microns}.  Colored symbols represent binned data that have been normalized with respect to the stellar flux (during secondary eclipse).  The solid black lines indicate the best-fit models.  Secondary eclipses occur at phases of 0.5 and 1.5; primary transits occur at a phase of 1.0 and are clipped to highlight the planetary emission.  Figure \ref{fig:transits} displays the full transits.  We do not fit the faded binned points near an orbital phase of 0.6 in the second 3.6~{\micron} visit; their discrepancy is likely the result of unmodeled instrumental or astrophysical red noise commonly seen in {\em Spitzer} time-series data sets \citep[e.g.][]{Stevenson2010, Blecic2013, Cubillos2013, HDL2014}.
}}
\end{figure*}

\subsection{4.5~$\mu$m Phase Curve}

We detect a strong asymmetry in the 4.5~{\micron} phase curve that is similar in shape to the {\em HST}/WFC3 band-integrated phase curve \citep[][see Figure \ref{fig:PhaseCurves}]{Stevenson2014c}.  The median phase curve maximum occurs $69{\pm}6$ minutes prior to the midpoint of secondary eclipse, which corresponds to a shift of $21.1{\pm}1.8\degrees$ East of the substellar point.  The median phase curve minimum occurs $22{\pm}9$ minutes after the primary transit midpoint, or $6.8{\pm}2.7\degrees$ West of the anti-stellar point.  Therefore, the maximum planetary emission occurs $0.421{\pm}0.009$ orbits after the observed minimum, which is consistent at the 1.5$\sigma$ level with our reported WFC3 difference \citep[$0.436{\pm}0.005$ orbits,][]{Stevenson2014c}.  This suggests that these two wavelength regions probe similar depths within WASP-43b's atmosphere.

In addition to the phase curves offset, we measure a peak-to-peak amplitude, $A$\sb{p2p}, of $0.399{\pm}0.014\%$ at 4.5~{\microns}.  Note that since the maximum flux occurs prior to secondary eclipse, it is physically plausible for $A$\sb{p2p} to exceed the secondary eclipse depth ($0.383{\pm}0.008\%$).

\subsection{3.6~$\mu$m Phase Curves}
\label{sec:3.6results}

When fit individually, the two 3.6~{\micron} phase curves exhibit contradictory shapes with conflicting nightside emission levels.  The first visit is consistent with being symmetric, albeit with large uncertainties, while the second exhibits measurable asymmetry.  Furthermore, we measure peak-to-peak amplitudes of $0.244{\pm}0.023\%$ and $0.338{\pm}0.011\%$, respectively, which corresponds to a difference of 3.7$\sigma$ (see Figure \ref{fig:PhaseCurves}).  The larger uncertainty in the first visit indicates that the nightside emission is at least partly degenerate with the position-dependent systematic (see Figure \ref{fig:pointingHist}).

As a test, we fit both visits simultaneously using a common set of shared phase curve parameters ($c_1$ - $c_4$).  If the phase curve parameters from the first visit are completely degenerate with the position-dependent systematics then the combined peak-to-peak amplitude should favor the best-fit value from the second visit.  However, the resulting amplitude ($0.313{\pm}0.010\%$) is consistent with the error-weighted average of the two individual measurements; therefore, this degeneracy does not explain the significantly smaller peak-to-peak amplitude for the first visit.  Additionally, we note that the reduction in the number of free parameters is not justified ($\Delta$BIC = 41) compared to our final fits in which we do not share phase curve parameters.

When we combine the nightside emission from the first visit at 3.6~{\microns} with the lack of emission from WFC3 and 4.5~{\microns}, our cloud-free atmospheric retrievals (see Section \ref{sec:atm}) obtain unphysical results that require invoking disequilibrium chemistry with unrealistic abundances or a thermal inversion with an extreme C/O ratio.  For this reason, we determine that the second visit more accurately reflects WASP-43b's typical nightside emission and, thus, adopt those data for further analysis and interpretation.  With that said, in Section \ref{sec:gcm} we briefly revisit this discrepancy by comparing both 3.6~{\micron} phase curves to the 3D general circulation model (GCM) results from \citet{Kataria2015}.

For the second 3.6~{\micron} visit, the median phase curve maximum and minimum occur $40{\pm}6$ minutes prior to secondary eclipse and  $28{\pm}17$ minutes after primary transit, respectively.  This corresponds to shifts of $12.2{\pm}1.7\degrees$ East and $9{\pm}5\degrees$ West of the substellar and anti-stellar points.  The maximum-to-minimum orbital phase difference of $0.442{\pm}0.016$ is consistent with both the WFC3 and 4.5~{\micron} values.

\subsection{Disk-Integrated Brightness Temperatures}

For comparison with \cite{Stevenson2014c}, we convert the measured planetary emission on the day and night sides to disk-integrated brightness temperatures, $T$\sb{DIB}.  At a phase of 0.5, we measure the dayside $T$\sb{DIB} to be $1624{\pm}23$K at 3.6~{\microns} and $1512{\pm}25$K at 4.5~{\microns}.  On the planet nightside, we place 2$\sigma$ upper limits of 720 and 650~K at 3.6 and 4.5, respectively.  The higher temperatures at 3.6~{\microns} suggest that this channel probes deeper within WASP-43b's atmosphere (assuming a non-inverted thermal profile).

\subsection{Transits and Eclipses}

\begin{figure}[tb]
\centering
\includegraphics[width=1.0\linewidth]{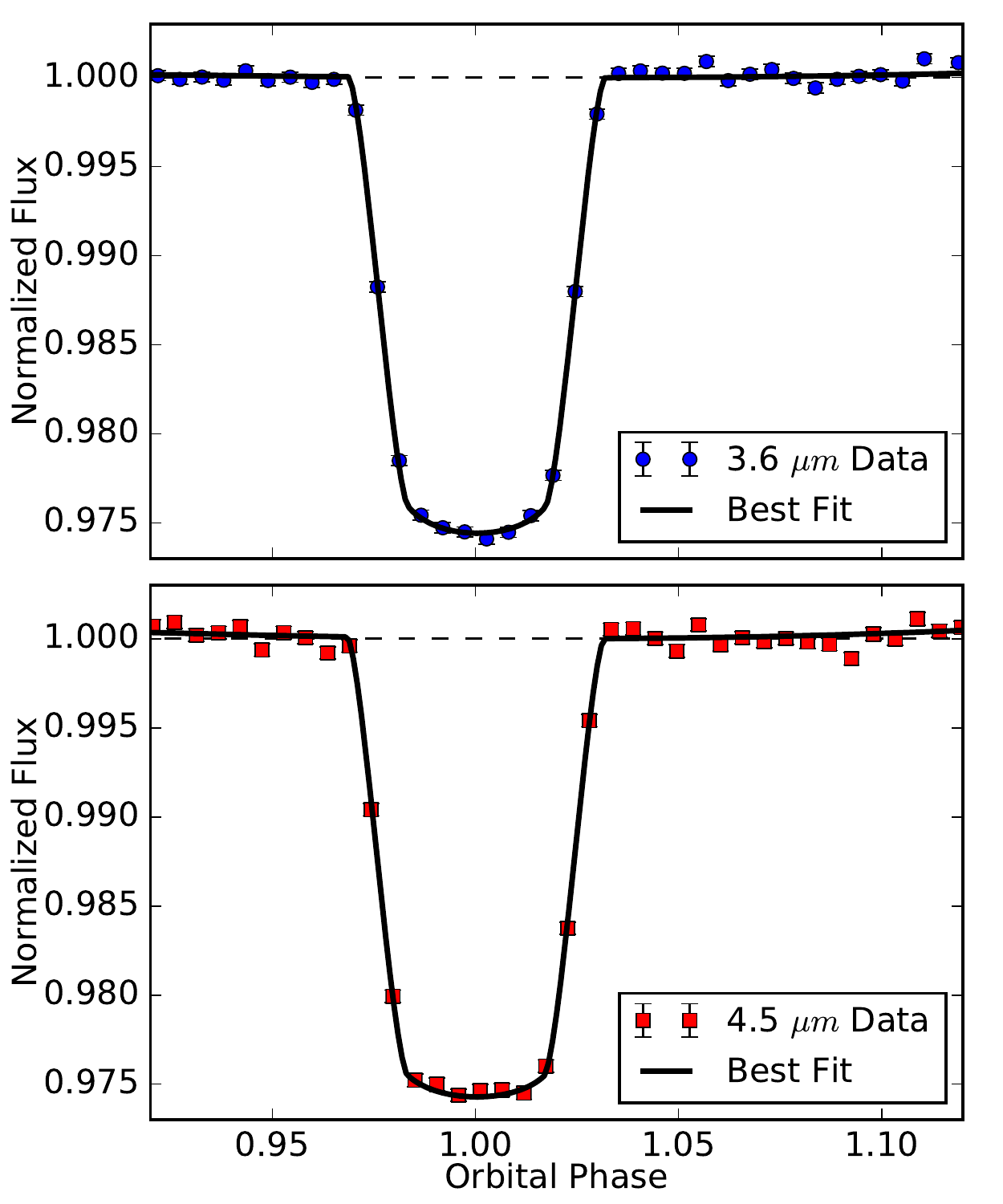}
\caption{\label{fig:transits}{
Primary transits of WASP-43b at 3.6 (top) and 4.5 (bottom)~{\microns}.  Colored symbols represent binned data that have been normalized with respect to the stellar flux.  The solid black lines indicate the best-fit models.
}}
\end{figure}

\begin{figure}[tb]
\centering
\includegraphics[width=1.0\linewidth]{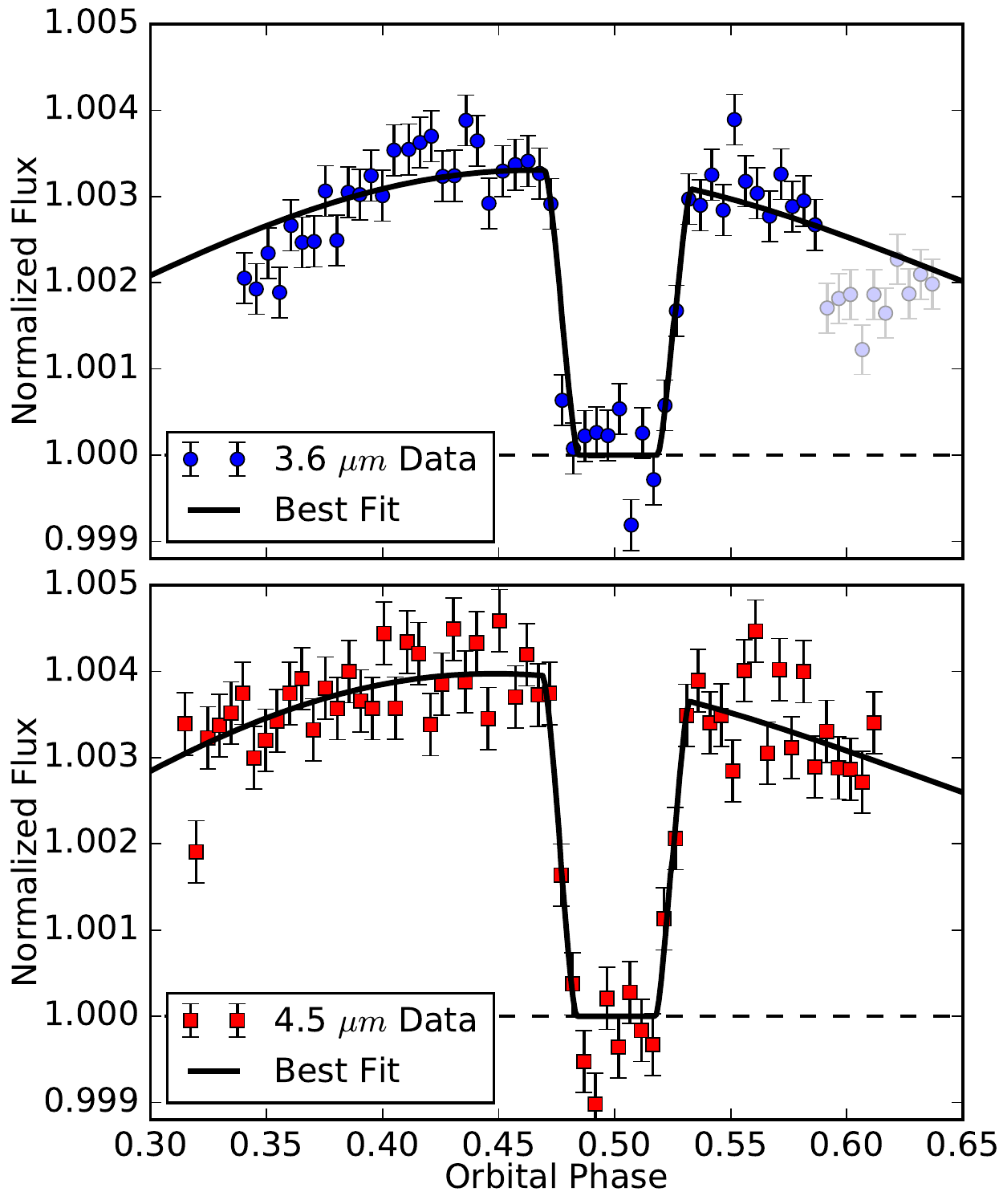}
\caption{\label{fig:eclipses}{
Secondary eclipses of WASP-43b at 3.6 (top) and 4.5 (bottom)~{\microns} from the 2011 observations.  Colored symbols represent binned data that have been normalized with respect to the stellar flux.  The solid black lines indicate the best-fit models.  We do not fit the faded binned points near the end of the 3.6~{\micron} dataset.
}}
\end{figure}

Each phase curve observation contains two secondary eclipses and one primary transit.  Figure \ref{fig:PhaseCurves} displays the former and Figure \ref{fig:transits} displays the latter.  As part of this study, we also reanalyzed the secondary eclipse data originally published by \citet{Blecic2014}.  In addition to fitting an eclipse model to these data, we include a model that fits the planet's phase-dependent flux variation (see Figure \ref{fig:eclipses}).  In our final analysis, we fit all datasets simultaneously and share common parameter values (i.e. eclipse depth, $c_1$ - $c_4$) between datasets.  Table \ref{tab:params} lists the best-fit parameters with 1$\sigma$ uncertainties from our joint fit.

\begin{table}[tb]
\centering
\caption{\label{tab:params} 
Best-Fit Parameters}
\begin{tabular}{lll}
    \hline
    \hline
    Parameter                   & 3.6~{\micron} Value\tablenotemark{a}  
                                                        & 4.5~{\micron} Value\tablenotemark{a} \\
    \hline
    Transit Times (BJD\sb{TDB}) & 2457089.11181(7)      & 2456897.13195(7)  \\
                                & 2457270.51672(6)      &                   \\
    $R\sb{p}/R\sb{\star}$       & 0.1580(3)             & 0.1589(5)         \\
    $a/R$\sb{$\star$}           & 4.855\tablenotemark{b}    & 4.855\tablenotemark{b}       \\
    $\cos i$                    & 0.13727\tablenotemark{b}  & 0.13727\tablenotemark{b}     \\
    Eclipse Times (BJD\sb{TDB}) & 2455773.3182(4)       & 2455772.5045(5)   \\
                                & 2457088.7048(6)       & 2456896.7256(4)   \\
                                & 2457089.5195(6)       & 2456897.5404(4)   \\
                                & 2457270.1109(5)       &                   \\
                                & 2457270.9235(4)       &                   \\
    Eclipse Depth (\%)          & 0.323(6)              & 0.383(8)          \\
    Eclipse Duration ($t$\sb{1-4}, days)
                                & 0.051300\tablenotemark{b}         & 0.051300\tablenotemark{b} \\
    Ingress/Egress ($t$\sb{1-2}, days)
                                & 0.011753\tablenotemark{b}         & 0.011753\tablenotemark{b} \\
    $c$\sb{1} (\%)              & 0.163(6)\tablenotemark{c}         & 0.193(6)\\
    $c$\sb{2} (BJD\sb{TDB})     & 2457088.690(4)\tablenotemark{c}   & 2456896.706(3)  \\
    $c$\sb{3} (\%)              & 0.025(4)\tablenotemark{c}         & 0.028(4)\\
    $c$\sb{4} (BJD\sb{TDB})     & 2457088.658(10)\tablenotemark{c}  & 2456896.622(10)  \\
    $\chi^2_{\nu}$              & 1.01                              & 1.25  \\
    \hline
\end{tabular}
\tablenotetext{1}{Parentheses indicate 1$\sigma$ uncertainties in the least significant digit(s).}
\tablenotetext{2}{Fixed to the best-fit value from \citet{Stevenson2014c}.}
\tablenotetext{3}{Values from second 3.6~{\micron} visit only.}
\end{table}

\begin{table}[tb]
\centering
\caption{\label{tab:depths} 
Individual Eclipse Depths}
\begin{tabular}{ccc}
    \hline
    \hline
    Label       & Wavelength    & Eclipse Depth\tablenotemark{a} \\ 
                & ({\microns})  & (\%)          \\
    \hline
    wa043bs11   & 3.6           & 0.356(13)     \\
    wa043bo11   & 3.6           & 0.324(16)     \\
                & 3.6           & 0.310(16)     \\
    wa043bo12   & 3.6           & 0.343(13)     \\
                & 3.6           & 0.306(12)     \\
    wa043bs21   & 4.5           & 0.412(14)     \\
    wa043bo21   & 4.5           & 0.365(14)     \\
                & 4.5           & 0.369(13)     \\
    \hline
\end{tabular}
\tablenotetext{1}{Parentheses indicate 1$\sigma$ uncertainties in the least significant digit(s).}
\end{table}

For our analyses, we test for signs of variability by measuring individual eclipse depths in a joint fit (see Table \ref{tab:depths}).  We note that the measured eclipse depths over different epochs and even between sequential visits vary more than expected given our computed uncertainties.  Nonetheless, all of the individual depths are within 2$\sigma$ of their shared, best-fit values.  Since the variations are not statistically significant, we conclude that using a single eclipse depth at each wavelength adequately represents the combined measurements.

We also compare our individually measured eclipse depths from the 2011 observations to those reported by \citet{Blecic2014}.  At both wavelengths, we achieve slightly deeper eclipse depths (by $<1.5\sigma$).  This may be because we crop the final 1,750 data points from our fits at 3.6~{\microns} and, at 4.5~{\microns}, our models include the phase curve variation whereas \citet{Blecic2014} do not apply a ramp model.

We determine the {\em Spitzer} transit depths to be $2.496{\pm}0.009\%$ at 3.6~{\microns} and $2.525{\pm}0.016\%$ at 4.5~{\microns}.  These values are noticeably shallower than the mean WFC3 transit depth \citep[2.5434\%,][]{Kreidberg2014b}.  Thus, the slope in the transmission spectrum ($\Delta Z$\sb{J-LM}$/H = 5.1{\pm}1.1$) suggests the presence of hazes that are partially obscuring the signal \citep{Sing2016}.  This is consistent with the need for a cloud deck in our WFC3 atmospheric retrieval \citep{Kreidberg2014b} and the interpretation of \citet{Stevenson2016b}, in which we find that the WFC3 water feature only extends $1.1{\pm}0.5$ planetary scale heights.  A cloud-free atmosphere should exhibit spectral features that extend over several scale heights.

\subsection{Orbital Constraints}

The short orbital period of WASP-43b suggests that star-planet tidal interactions are likely causing the planet's orbit to slowly decay.  Thus, one day WASP-43b could spiral into its host star.  The process of tidal decay manifests by a change in orbital period and may be observable over long baselines.  Previous constraints of orbital decay \citep{Blecic2014, Murgas2014} hinted at low-significance detections, but were dependent on less-precise ground-based measurements.  With the precision of {\em Spitzer} and its extended time baseline, we are in a position to better evaluate WASP-43b's rate of orbital decay.

\begin{figure}[tb]
\centering
\includegraphics[width=1.0\linewidth]{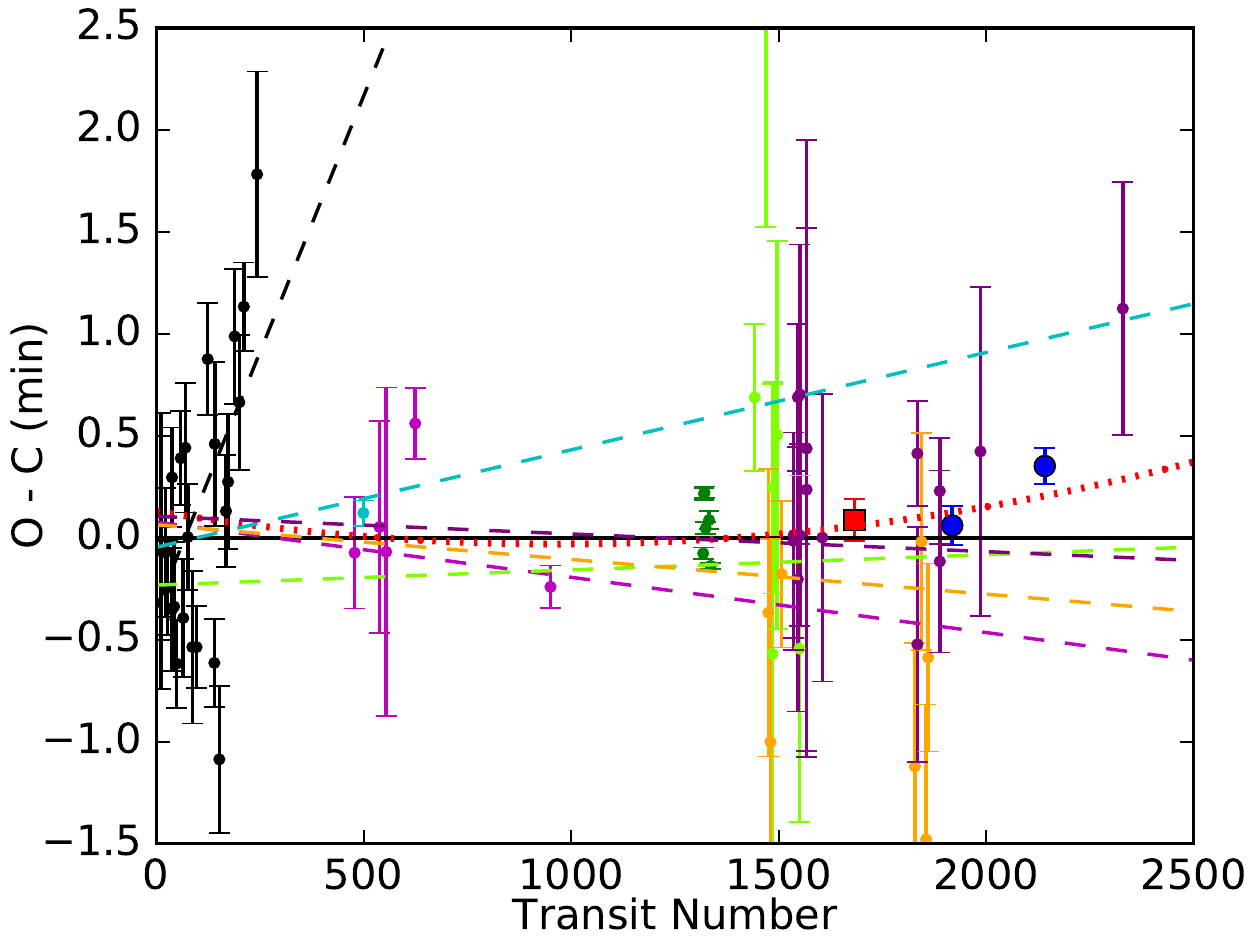}
\caption{\label{fig:orbit}{
Observed minus calculated ($O-C$) transit times of WASP-43b.
Colored points and dashed lines represent measured transit times and best-fit linear solutions from \citet[][black]{Gillon2012}, \citet[][cyan]{Chen2014}, \citet[][magenta]{Murgas2014}, \citet[][green]{Stevenson2014c}, \citet[][chartreuse]{Ricci2015}, \citet[][orange]{Jiang2016}, and \citet[][purple]{Hoyer2016}.  Including the transit times from this work (blue and red symbols), we find no evidence for tidal decay as illustrated by the positive trend in the best quadratic fit (dotted red line).
}}
\end{figure}

First, we apply a linear ephemeris model ($T_c = T_0 + N p$) to the observed transit times from \citet{Gillon2012}, \citet{Chen2014}, \citet{Murgas2014}, \citet{Stevenson2014c}, \citet{Ricci2015}, \citet{Jiang2016}, and \citet{Hoyer2016}.  By minimizing the error-weighted residuals in the observed minus calculated transit times, we compute a new ephemeris, $T_0 = 2455528.86856(3)$ BJD\sb{TDB}, and orbital period, $p = 0.81347403(3)$ days.  Next, we estimate the decay rate by adding a quadratic term to our ephemeris model \citep{Adams2010,Blecic2014}; however, we find a slight increase in the orbital period, $\dot{p} = 0.009{\pm}0.004$~s~yr\sp{-1}.  This can be seen in Figure \ref{fig:orbit}, where the shape of the best-fit quadratic solution is influenced by our most recent 3.6~{\micron} transit time.  The large spread in $O-C$ times may be due to underestimated uncertainties in individual measurements, transit timing variations, or stellar activity affecting the apparent transit times.  In conclusion, using transit times that span more than four years, we find no significant evidence for tidal decay in the orbit of WASP-43b.

\section{ATMOSPHERIC COMPOSITION}
\label{sec:atm}

\subsection{Achieving Independent Phase Curve Uncertainties}
\label{sec:ind}

In order to perform independent atmospheric retrievals at all orbital phases, we need phase-independent planetary emission uncertainty estimates.  The posterior distributions of the phase curve models (e.g., the colored regions in Figure~\ref{fig:gcm}) depict our knowledge of WASP-43b's emission at any given orbital phase; however, this information cannot be used to generate independent phase curve uncertainties for our 15 bins because the results would be highly correlated.  The standard error in the flux at each binned phase is a better choice, but it underestimates the true uncertainty in our measurements because it omits the absolute uncertainty in each channel.  

By choosing the in-eclipse flux as our baseline, we add (in quadrature) the secondary eclipse uncertainties and the standard errors at each binned phase to derive our final planetary emission uncertainties (see Table \ref{tab:spectra}).  We use the secondary eclipse uncertainty over the standard error during full eclipse because the ingress and egress durations are fixed; therefore, the additional data from the former provide a slightly more precise constraint.  The new strategy outlined here is opposite to the standard practice of using the out-of-eclipse data as baseline and allows us to infer composition constraints using data from all phases except during secondary eclipse.

\begin{figure*}[t]
\centering
\includegraphics[width=1.0\linewidth]{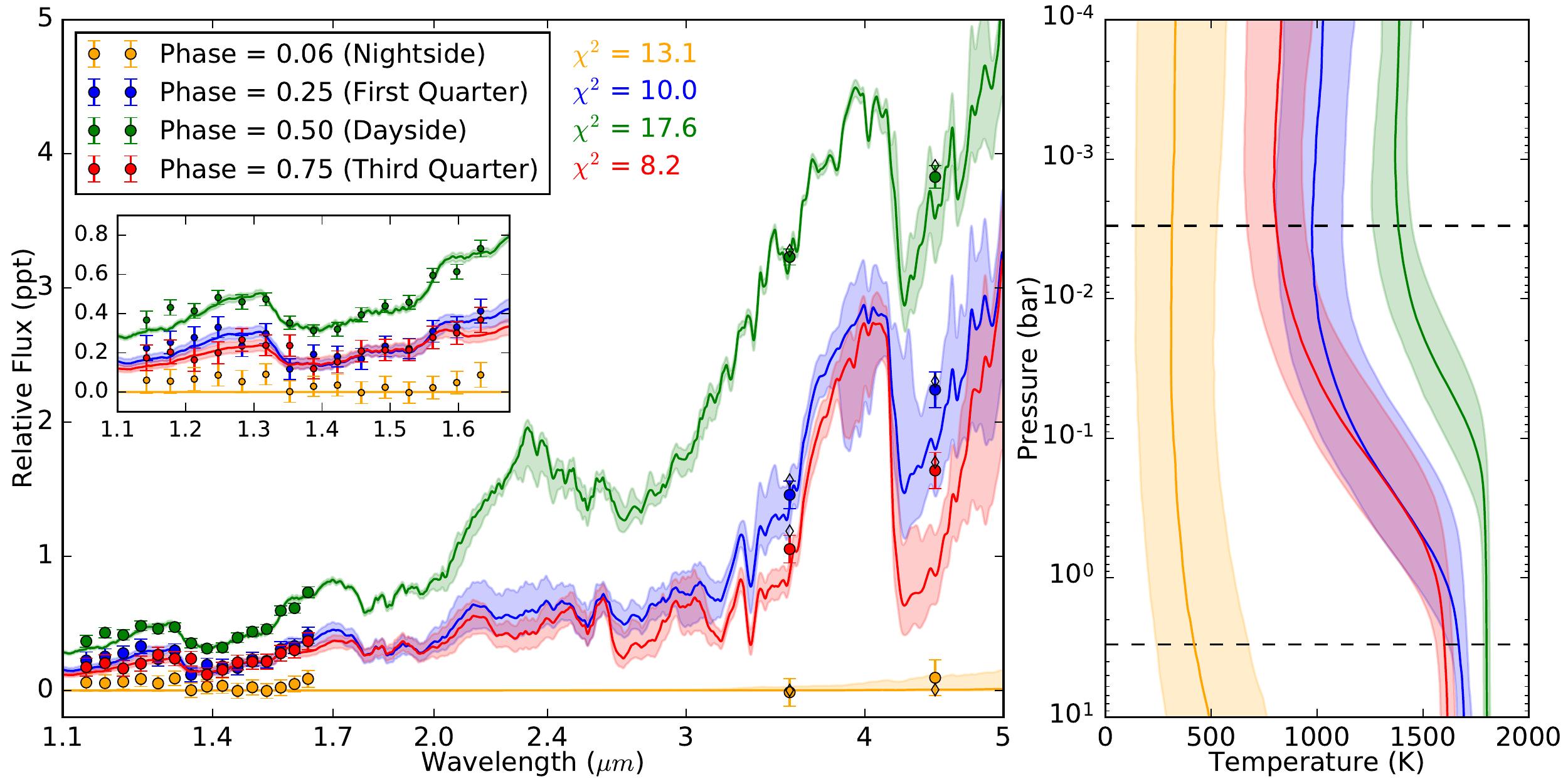}
\caption{\label{fig:spec}{
Emission spectra (left) and thermal profiles (right) of WASP-43b at four orbital phases.  Each set of 17 colored circles depict measurements from our {\em HST} and {\em Spitzer} phase curve observations. Colored curves with shaded regions represent median models with 1$\sigma$ uncertainties and are consistent with best-fit models at the 0.4$\sigma$ level.  Diamonds depict the {\em Spitzer} bandpass-integrated models, most of which overlap the measured values.  The inset magnifies the WFC3 spectra.  For comparison, the nightside data (phase = 0.06) from the first 3.6~{\micron} visit has a measured relative flux of 1.51{\pm}0.08 ppt.  The dashed horizontal lines in the right panel depict the pressure limits of our contribution functions; our thermal profiles are valid only between these lines.
}}
\end{figure*}

\subsection{Atmospheric Retrieval Models}

We derive the planet's atmospheric composition and thermal structure using the CHIMERA Bayesian retrieval suite, which is described in detail by \citet{Line2013a, Line2014-C/O} and was previously used to interpret WASP-43b's atmosphere using {\em HST}/WFC3 data \citep{Kreidberg2014b, Stevenson2014c}.  As with previous work, we adopt a 1D thermal profile representing the hemispherical-average temperature structure at each orbital phase.  We retrieve abundances (volume mixing ratios) for six prominent molecules (H\sb{2}O, CH\sb{4}, CO, CO\sb{2}, HCN, and NH\sb{3}), some of which are poorly constrained because they are not thermochemically favored on the dayside of WASP-43b.  We use Phoenix stellar grid models \citep{Allard2000} interpolated to $\log g = 4.646$ and $T_{eff} = 4400$~K.

We perform independent atmospheric retrievals to the binned {\em HST} and {\em Spitzer} light-curve data (excluding the first 3.6~{\micron} visit) at 15 orbital phases.  We then repeat the process using the best-fit light-curve models evaluated at the same orbital phases and using the independent uncertainties discussed in Section \ref{sec:ind}.  The two methods produce comparable results at most orbital phases, but the latter is less susceptible to hour-long-scale red noise in the {\em Spitzer} light curves; therefore, we present those results in the discussion and figures below.

In Figure \ref{fig:spec}, we depict measured {\em HST} and {\em Spitzer} emission spectra with median model spectra and thermal profiles at four complementary orbital phases.  The median fits are consistent with our best-fit solutions at the 0.4$\sigma$ level.  We achieve good fits at all orbital phases ($\chi^2= 7.0 - 17.6$, 17 data points) and comparable results at first and third quarters.  The models at these two phases suggest the presence of strong features in the {\em Spitzer} bandpasses that are sculpted by the absorption of CH\sb{4} near 3.3~{\microns} and CO/CO\sb{2} near 4.4~{\microns}.  However, by employing two 1D thermal profiles, \citet{Feng2016} reduce the amount of CH\sb{4} needed to achieve a good fit, thus decreasing the peak-to-trough amplitude at first and third quarters.  We discuss the effects of adding a second thermal profile in more detail below, but to summarize our findings, the large variations in the first- and third-quarter models depicted in Figure \ref{fig:spec} are unlikely to represent WASP-43b's true emission spectra at those orbital phases.

\subsection{Cloud-Free Molecular Abundances}
\label{sec:abundances}

\begin{figure*}[t]
\centering
\includegraphics[width=0.49\linewidth]{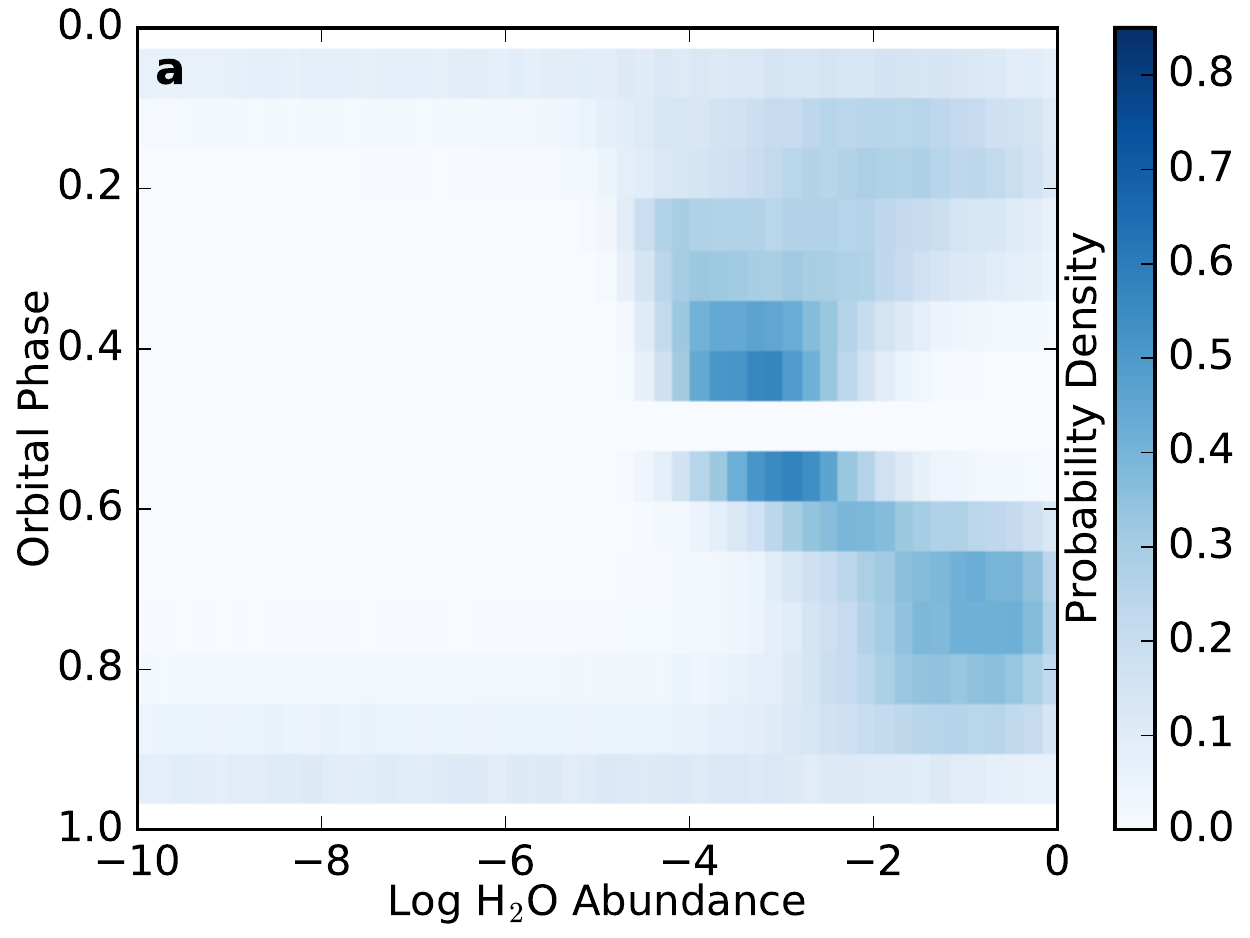}
\includegraphics[width=0.49\linewidth]{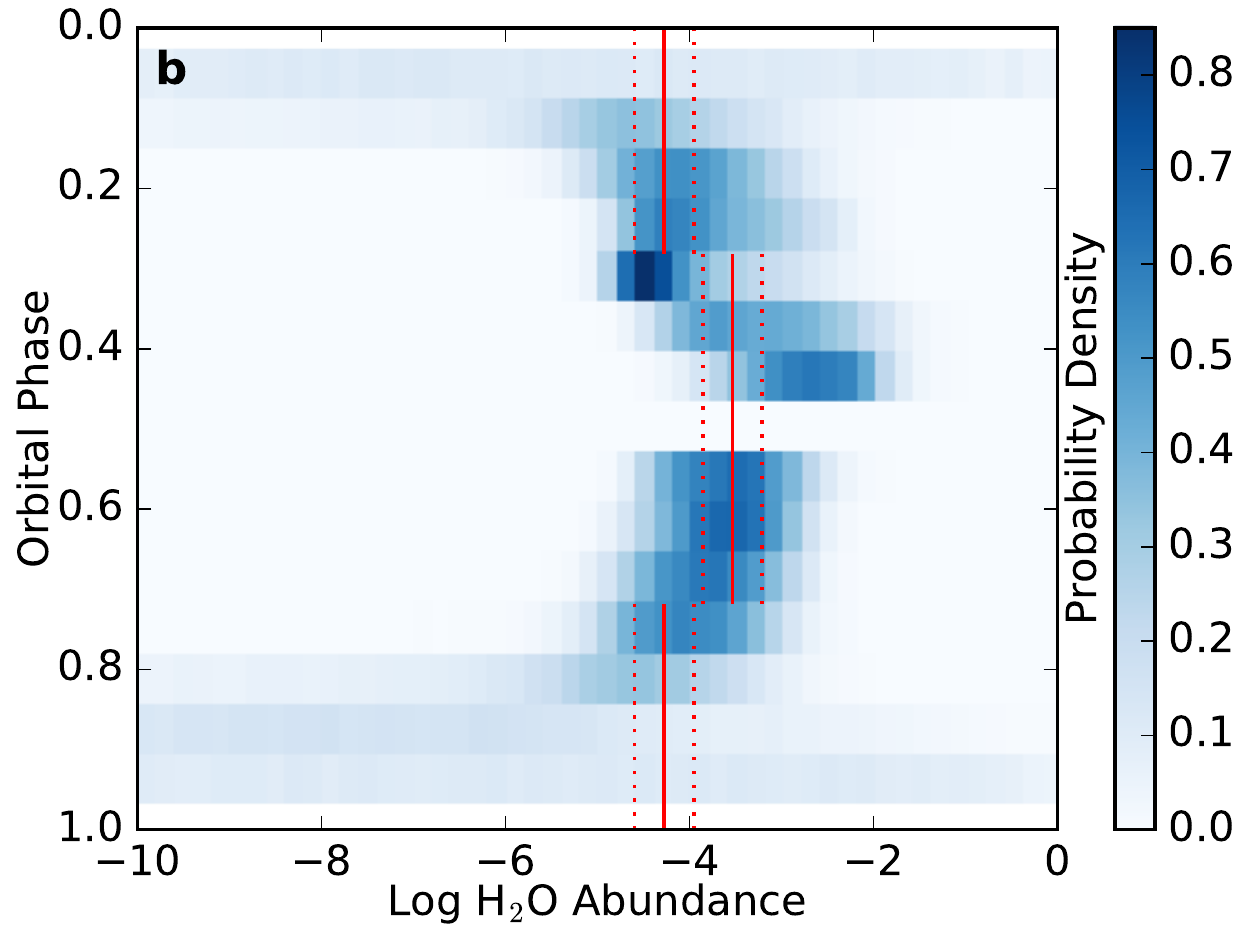}
\includegraphics[width=0.49\linewidth]{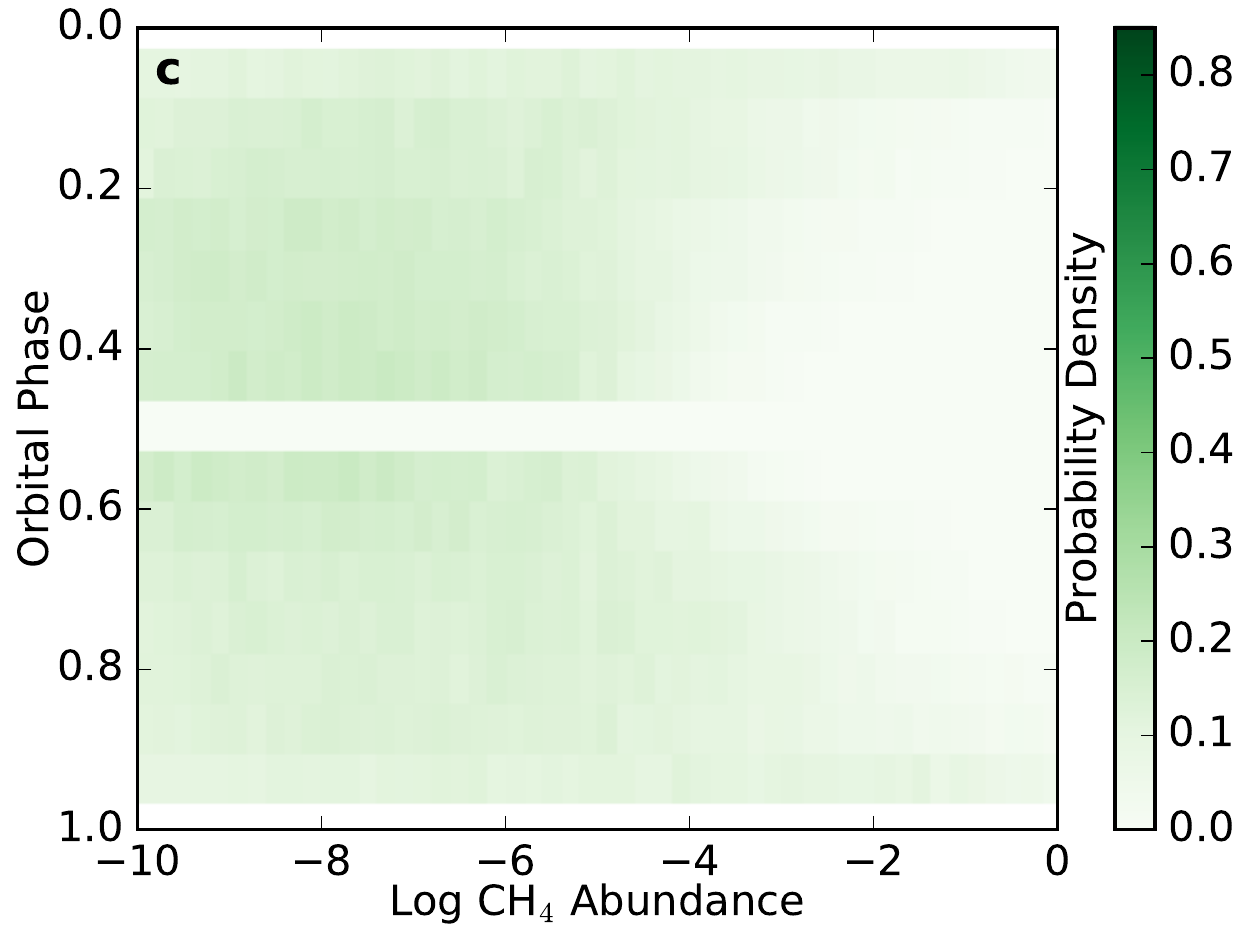}
\includegraphics[width=0.49\linewidth]{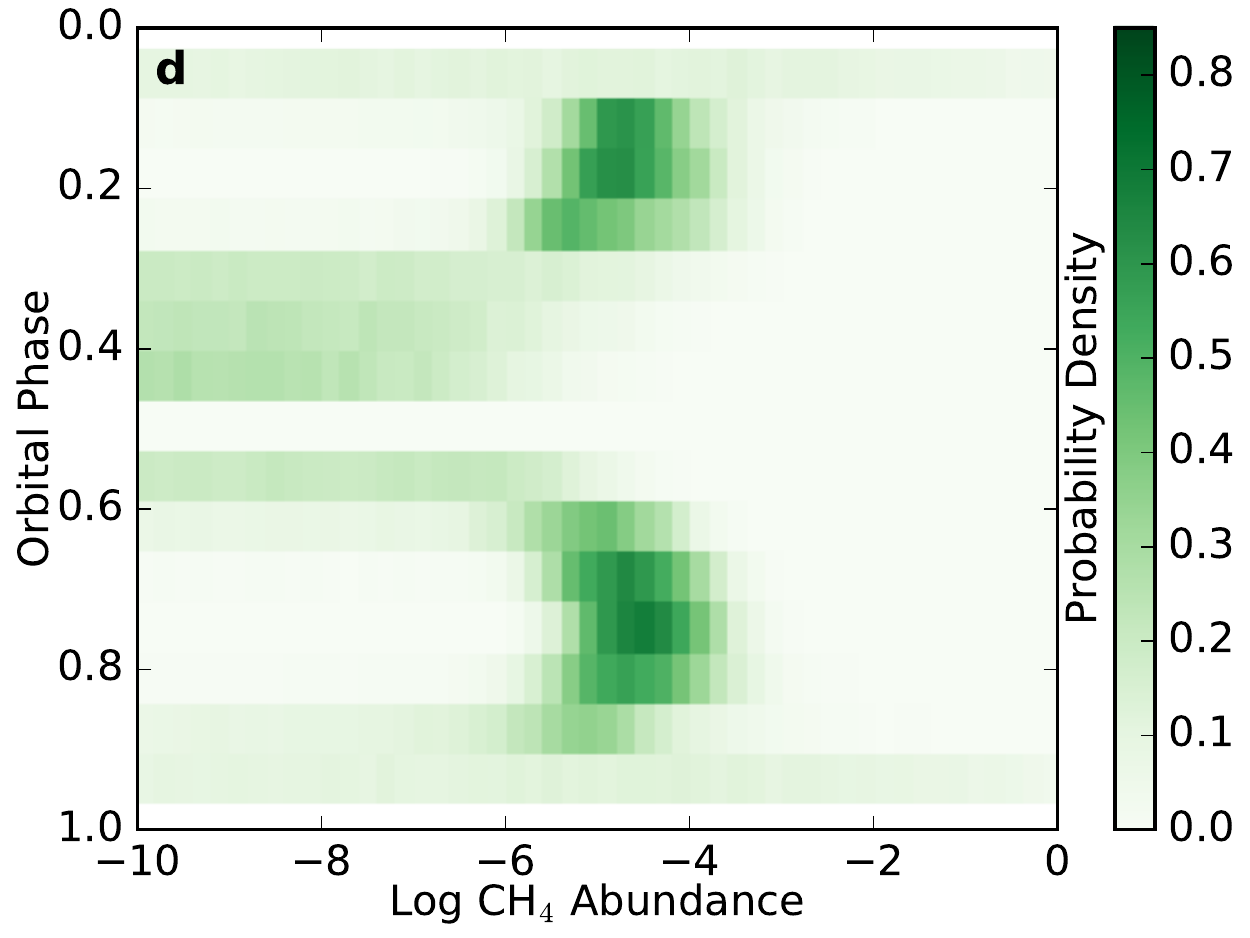}
\includegraphics[width=0.49\linewidth]{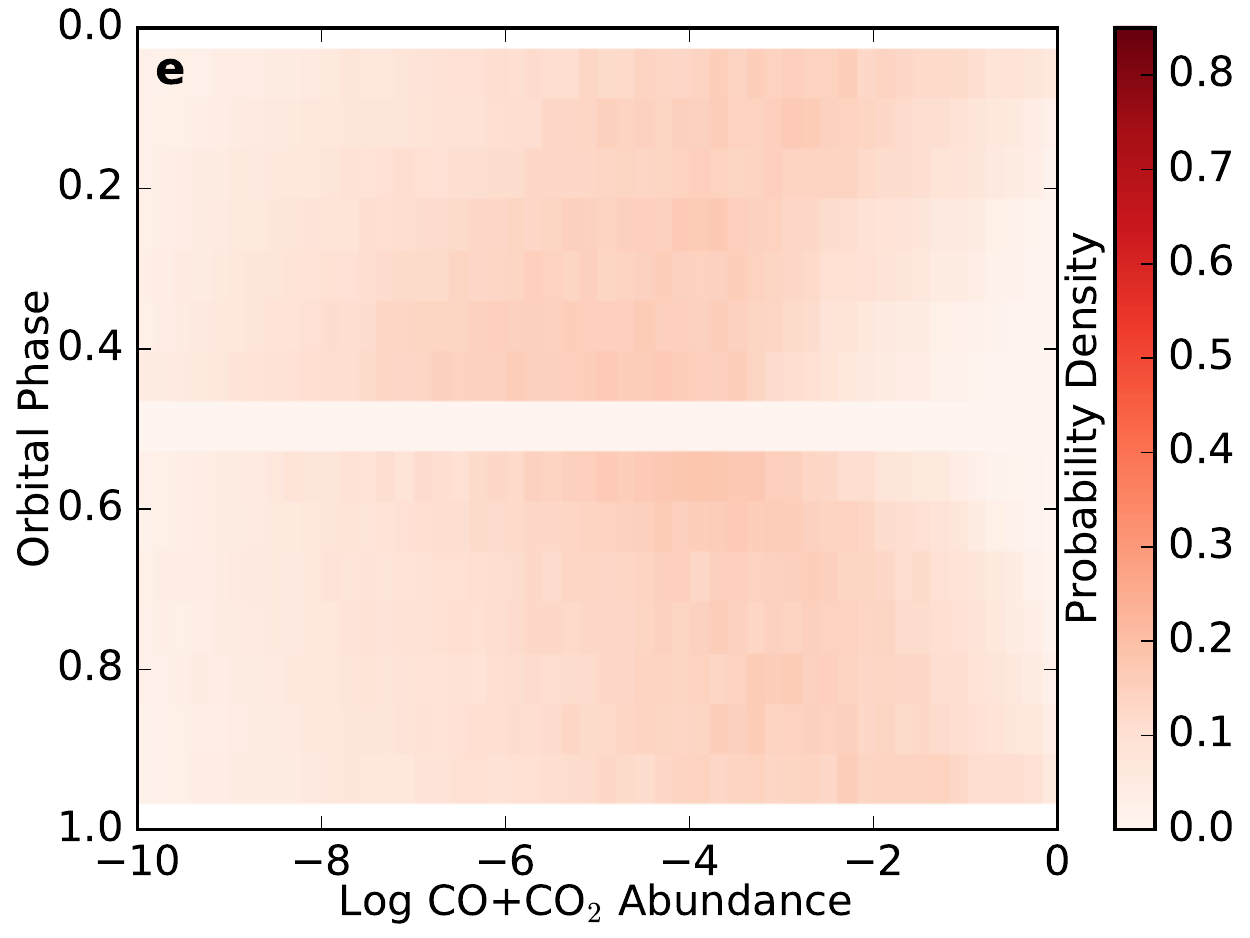}
\includegraphics[width=0.49\linewidth]{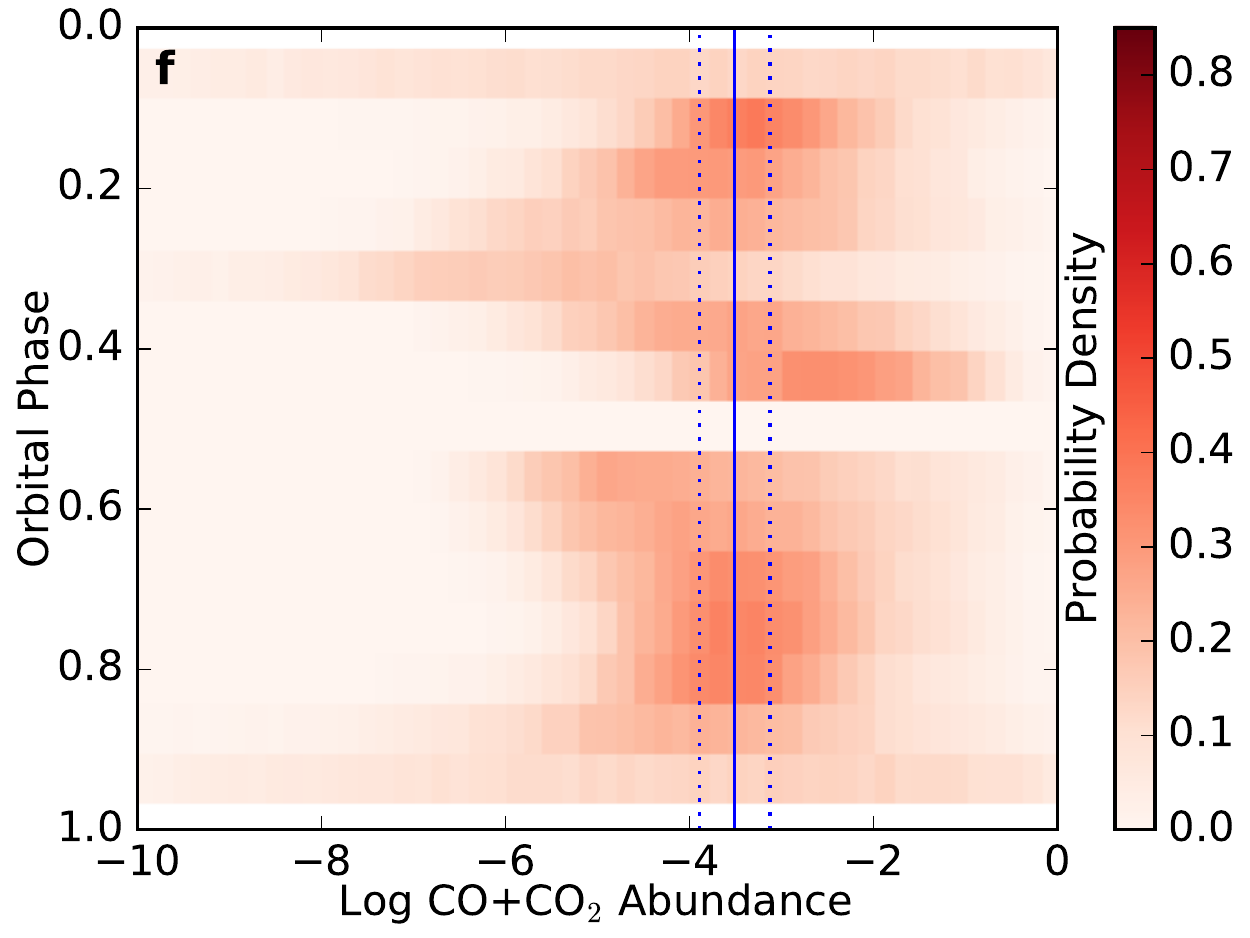}
\caption{\label{fig:abundances}{
Log molecular abundance constraints for WASP-43b using the WFC3 data only (left column) and WFC3+IRAC data (right column).  The colored histograms depict probability density regions computed independently at each orbit phase.  Vertical solid and dotted lines represent mean abundances with {\pm}1$\sigma$ uncertainty regions over the indicated orbital phases.  In panel (d), the bounded CH\sb{4} constraints near first and third quarters likely result from using individual, average thermal profiles to represent contributions from both a hot dayside and a cold nightside \citep{Feng2016}.  As discussed in Section \ref{sec:ind}, the secondary eclipse bin contains no independent information and, thus, is not shown in any of the panels.
}}
\end{figure*}

In Figure \ref{fig:abundances}, we compare H\sb{2}O, CH\sb{4}, and CO+CO\sb{2} abundance constraints both with and without the {\em Spitzer}/IRAC data.  The abundances of CO and CO\sb{2} are degenerate because both molecules have absorption features in {\em Spitzer}'s 4.5~{\microns} bandpass that cannot be individually resolved; therefore, we combine these two molecules into a single constraint.  Using only the {\em HST}/WFC3 data, we obtain a bounded constraint on the abundance of H\sb{2}O at most orbital phases and only an upper limit on the abundances of CH\sb{4} and CO+CO\sb{2}.  This is to be expected since the WFC3 bandpass contains an absorption feature for water (but not carbon monoxide or carbon dioxide) and methane is not predicted to form in any appreciable amount at the temperatures exhibited on WASP-43b's dayside.  Due to horizontal quenching \citep{Cooper2006, Agundez2014}, we expect constant dayside abundances at all orbital phases.

\begin{figure}[t]
\centering
\includegraphics[width=1.0\linewidth]{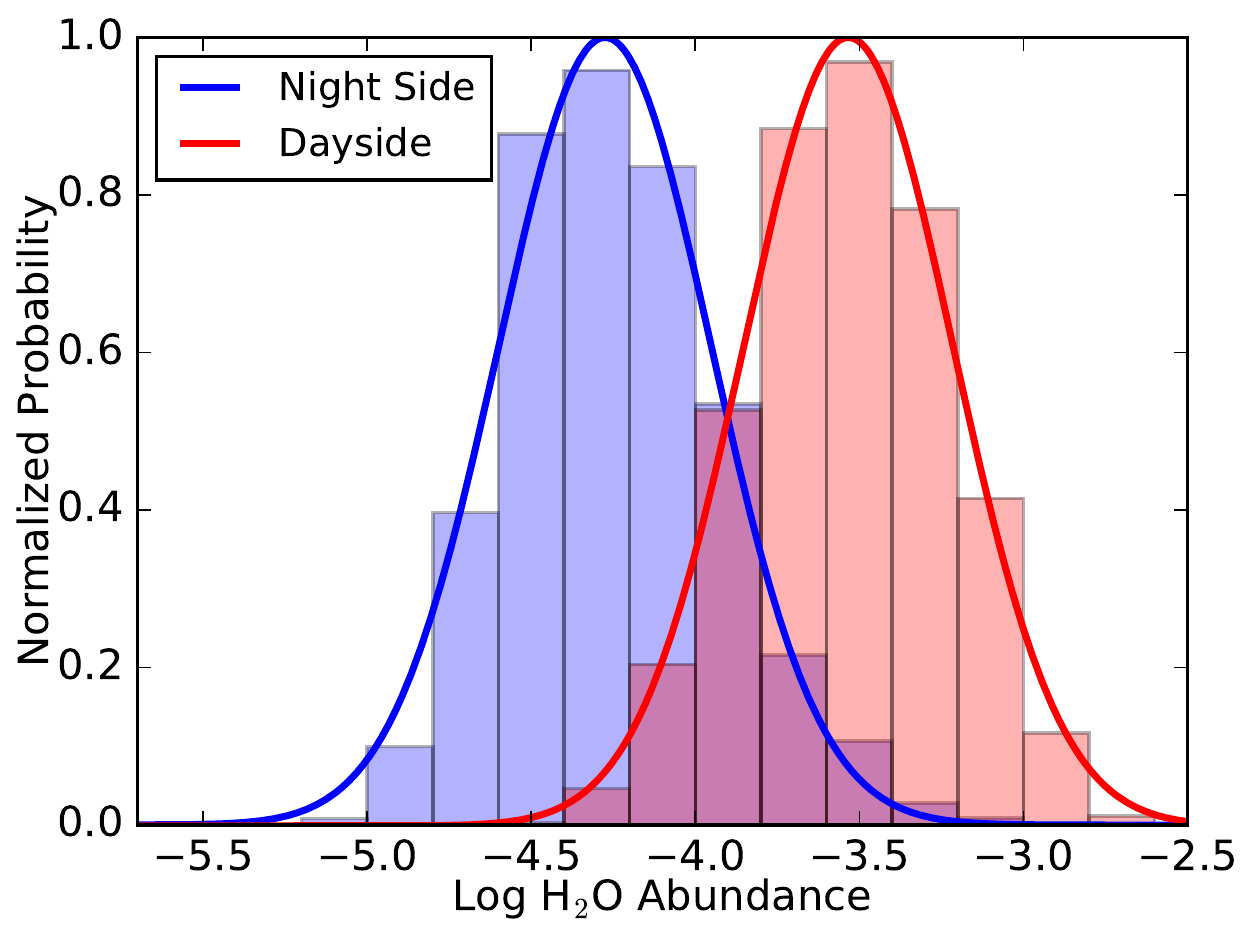}
\includegraphics[width=1.0\linewidth]{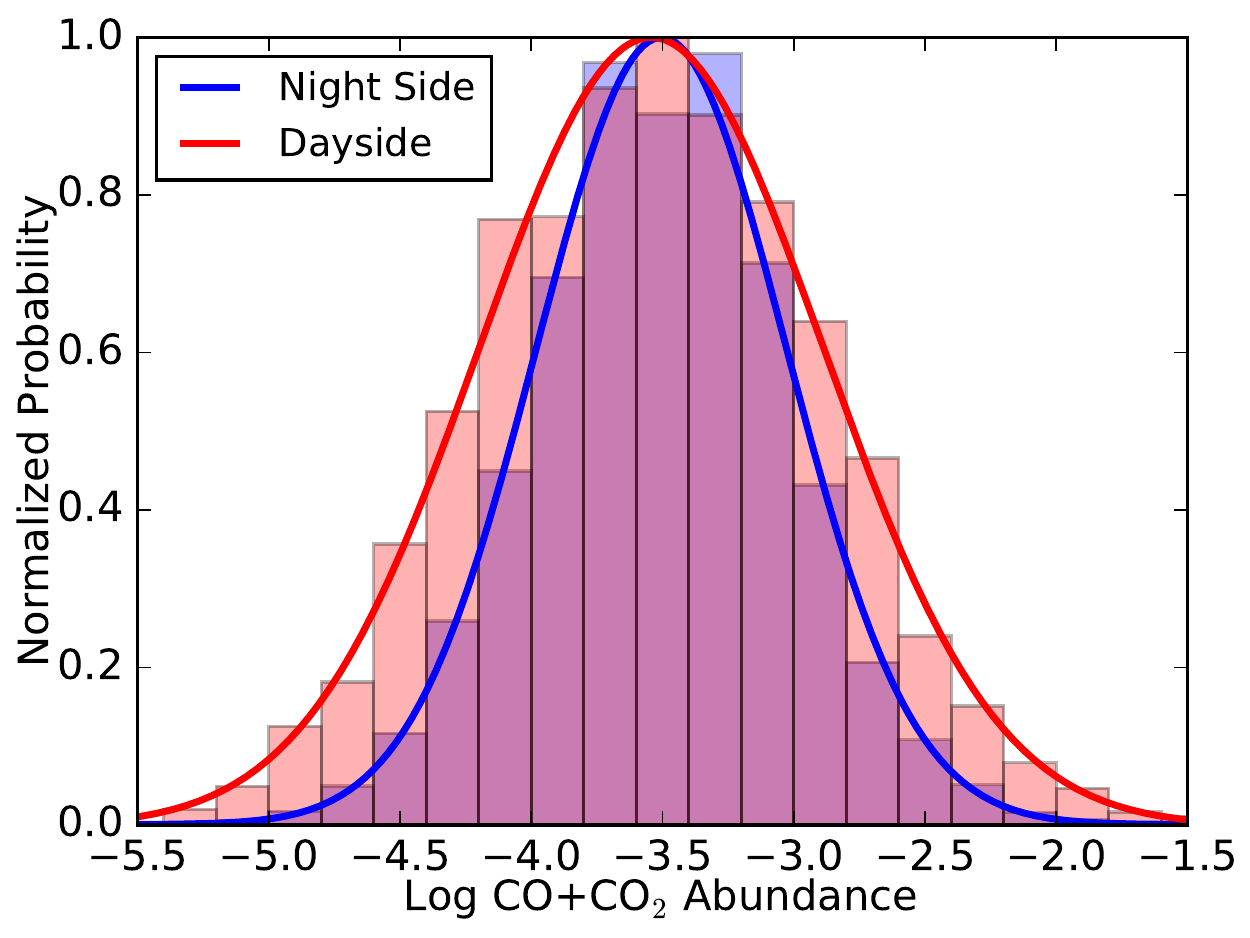}
\caption{\label{fig:daynightabundance}{
Log molecular abundance constraints for H\sb{2}O (top) and CO+CO\sb{2} (bottom) when considering the planet dayside (red) and nightside (blue).  The measured abundances of CO+CO\sb{2} are consistent (within our retrieval uncertainties) across both hemispheres.  Our retrievals suggest that the H\sb{2}O abundance varies with orbital phase, but additional work is needed to investigate and eliminate any potential biases within our models.  The purple region indicates where the two histograms overlap.
}}
\end{figure}

When we include information from the {\em Spitzer} phase curves (see Figure \ref{fig:abundances}, right panels), we obtain bounded constraints on the H\sb{2}O and CO+CO\sb{2} abundances at all orbital phases for which we detect planetary emission.  To test whether these abundances are independent of orbital phase, we divide the distributions into two groups: ``dayside hemisphere'' (orbital phase = 0.28 $\to$ 0.72) and ``nightside hemisphere'' (0.72 $\to$ 0.28).  We then compute the product of the distributions within each group and fit a Gaussian function to determine a mean abundance and standard deviation (see Figure \ref{fig:daynightabundance}).  Performing independent two-sample $t$-tests and computing $p$-values for each molecule \citep{Bevington2003}, we find that the day-night abundance difference is statistically significant for H\sb{2}O ($p\sim0.001$) but not so for CO+CO\sb{2} ($p\sim0.9$).  

The use of a single, average thermal profile between a hot dayside and a cold nightside could bias the retrieved H\sb{2}O abundances away from the planet dayside; however, \citet{Feng2016} show that the H\sb{2}O abundance does not change significantly with the addition of a second thermal profile.  Furthermore, chemical equilibrium predicts that the abundance of H\sb{2}O should increase with decreasing temperature, as in panel (a) of Figure \ref{fig:abundances}, but we see the opposite trend in panel (b).  Thus, the observed phase-dependence in the water abundance could be due to an unidentified bias within our retrievals.  This result motivates new work to explore the complexities of atmospheric retrieval along the lines of \citet{Feng2016}.

Assuming the dayside- and nightside-hemisphere H\sb{2}O abundances are different, we determine $\log$ abundances of $-3.5{\pm}0.3$ and $-4.3{\pm}0.3$, respectively.  As a test, we perform an atmospheric retrieval on the error-weighted mean spectrum of the dayside hemisphere and determine a $\log$ H\sb{2}O abundance of $-3.9{\pm}0.2$.  To estimate the mean CO+CO\sb{2} abundance, we compute the product of 14 probability densities (excluding secondary eclipse) and fit a Gaussian to the resulting distribution.  WASP-43b's global $\log$ abundance of CO+CO\sb{2} is $-3.5{\pm}0.4$.

As seen in panel (d) of Figure \ref{fig:abundances}, the CH\sb{4} abundance appears to vary with orbital phase.  This goes against the expectation of a constant CH\sb{4} abundance due to horizontal quenching.  \citet{Feng2016} show that the bounded CH\sb{4} constraint near first quarter is likely an artificial byproduct of adopting a single thermal profile to represent contributions from both a hot dayside and a cold nightside.  When they add a second thermal profile to the retrieval, the CH\sb{4} abundance goes from a bounded constraint to an upper limit that is consistent with that from the planet dayside.  We estimate the dayside 2$\sigma$ upper limit on the $\log$ abundance to be -5.3.  Thus, using two thermal profiles at all orbital phases (with the proper weighting), we should expect an unbiased CH\sb{4} abundance that is constant with orbital phase.

We conclude that the bounded methane constraint near first and third quarters is driven by the retrieval trying to fit {\em Spitzer}'s 3.6~{\micron} point using only a single thermal profile, whereas the water abundance is determined primarily by the WFC3 data.  The latter measurements probe deeper within the planet's atmosphere where temperatures are expected to exhibit a smaller day-night contrast and, thus, would be less dependent on the number of thermal profiles in our model.  An explanation for the strong difference in retrieved day-night H\sb{2}O abundances could be the presence of high-altitude, obscuring clouds on the planet nightside.  We discuss this possibility below; however, without definitive evidence for nightside clouds, we recommend adopting the dayside hemisphere value where the measurements and retrievals are more robust.

\subsection{Molecular Abundances With Clouds}
\label{sec:abundances2}

To try to explain the retrieved phase-dependent H\sb{2}O abundances, we perform a test that includes clouds in our nightside and first quarter retrievals.  Our cloud model is parameterized with a cloud base pressure, scale height, and gray opacity \citep{Line2016a,Line2016b}.  Upon comparing the results, for which the molecular abundances and thermal profiles remain consistent at 1$\sigma$, we find no strong evidence to justify the inclusion of clouds in our models.  In fact, there is evidence against the cloud model (negative $\log$ Bayes factor) due to the increase in prior volume from the inclusion of three additional parameters without any accompanying improvement in fit.  Furthermore, the integrated column optical depth is much less than unity.  Simply put, the current data do not support this particular cloud model.  With additional observations from {\em JWST}, it is possible that this and more sophisticated models (such as non-uniform cloud cover models, whereby clouds only persist on the nightside) may be favored by the data.  

\subsection{Metallicity}

\begin{figure}[t]
\centering
\includegraphics[width=1.0\linewidth]{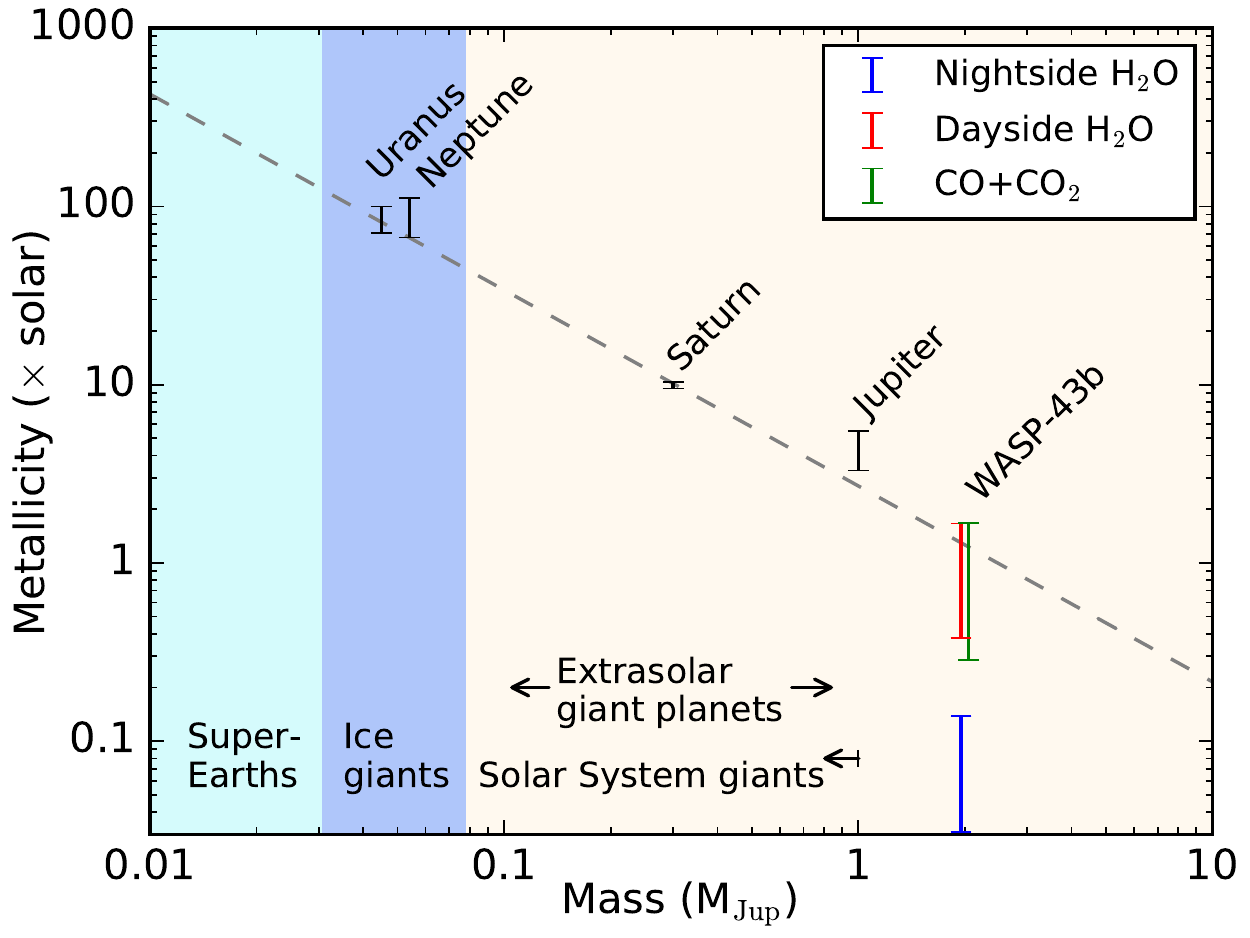}
\caption{\label{fig:metallicity}{
Updated WASP-43b atmospheric metallicity abundance compared to the solar system giant planets \citep[originally from][]{Kreidberg2014b}. 
The dashed line is an error-weighted power law fit to the data (1$\sigma$ uncertainties shown).  We infer the solar system planet metallicities from their measured methane abundances and WASP-43b's metallicity from our derived nightside water (blue), dayside water (red), and global CO+CO\sb{2} (green) abundances.  This figure further discredits the reliability of the retrieved nightside H\sb{2}O abundance.  These constraints utilize phase curve and secondary eclipse data from {\em HST}/WFC3 and {\em Spitzer}/IRAC (specifically the second 3.6~{\micron} visit and 4.5~{\microns} visit).
}}
\end{figure}

In \citet{Kreidberg2014b}, we used {\em HST}/WFC3 transit and eclipse observations to constrain the metallicity of WASP-43b (0.4 -- 3.5$\times$ solar).  Here, we refine the metallicity constraint by first multiplying the dayside- and nightside-hemisphere probability densities shown in panel (b) of Figure \ref{fig:abundances} to obtain two H\sb{2}O abundance constraints ($1.4\tttt{-4} - 6.1\tttt{-4}$ and $2.5\tttt{-5} - 1.1\tttt{-4}$ at 1$\sigma$ confidence).  We then convert these ranges to metallicities using temperature-dependent solar water volume mixing ratios at 0.1~bar \citep[$3.68\tttt{-4}$ at 1700~K and $8.13\tttt{-4}$ at 400~K,][]{Gordon1994,Lodders2002}.  Adopting the more reliable dayside hemisphere value, we find that WASP-43b's atmosphere is most likely solar in composition (0.4 -- 1.7 at 1$\sigma$ confidence).  This range is consistent with our previous estimate from \citet{Kreidberg2014b} and the trend exhibited by the solar system giant planets, as shown in Figure \ref{fig:metallicity}, but more precise because we include data from more orbital phases (0.28 $\to$ 0.72, excluding secondary eclipse).  Repeating the above calculation using the binned light-curve data (instead of evaluating the best-fit model), we find a similar range of dayside hemisphere metallicities.  We leave our investigation of the low nightside metallicity constraint for future work.  Using the global CO+CO\sb{2} abundance and a total volume mixing ratio of $4.5\tttt{-4}$ at 1700~K, we derive a markedly consistent metallicity constraint of 0.3 -- 1.7.

\section{ATMOSPHERIC CIRCULATION}
\label{sec:circ}

\subsection{Energy Budget}

By integrating the retrieved model spectra, we recompute dayside and nightside bolometric flux values and solve for the heat redistribution factor, $\mathcal{F} = 0.501^{+0.005}_{-0.001}$ \citep[where $\mathcal{F} = 0.5 \rightarrow 1$ spans the range from zero to full heat redistribution, derivation by ][]{Stevenson2014c}.  This value is consistent with our previous estimate ($\mathcal{F} = 0.503^{+0.021}_{-0.003}$), but more precise because of the {\em Spitzer} phase curve data.  We also re-derive a more precise estimate of the planet's Bond albedo ($A$\sb{b}$ = 0.19^{+0.08}_{-0.09}$) that is consistent with the expectation that hot Jupiters are generally black \citep{Fortney2008, Burrows2008}.

\begin{figure}[t]
\centering
\includegraphics[width=1.0\linewidth]{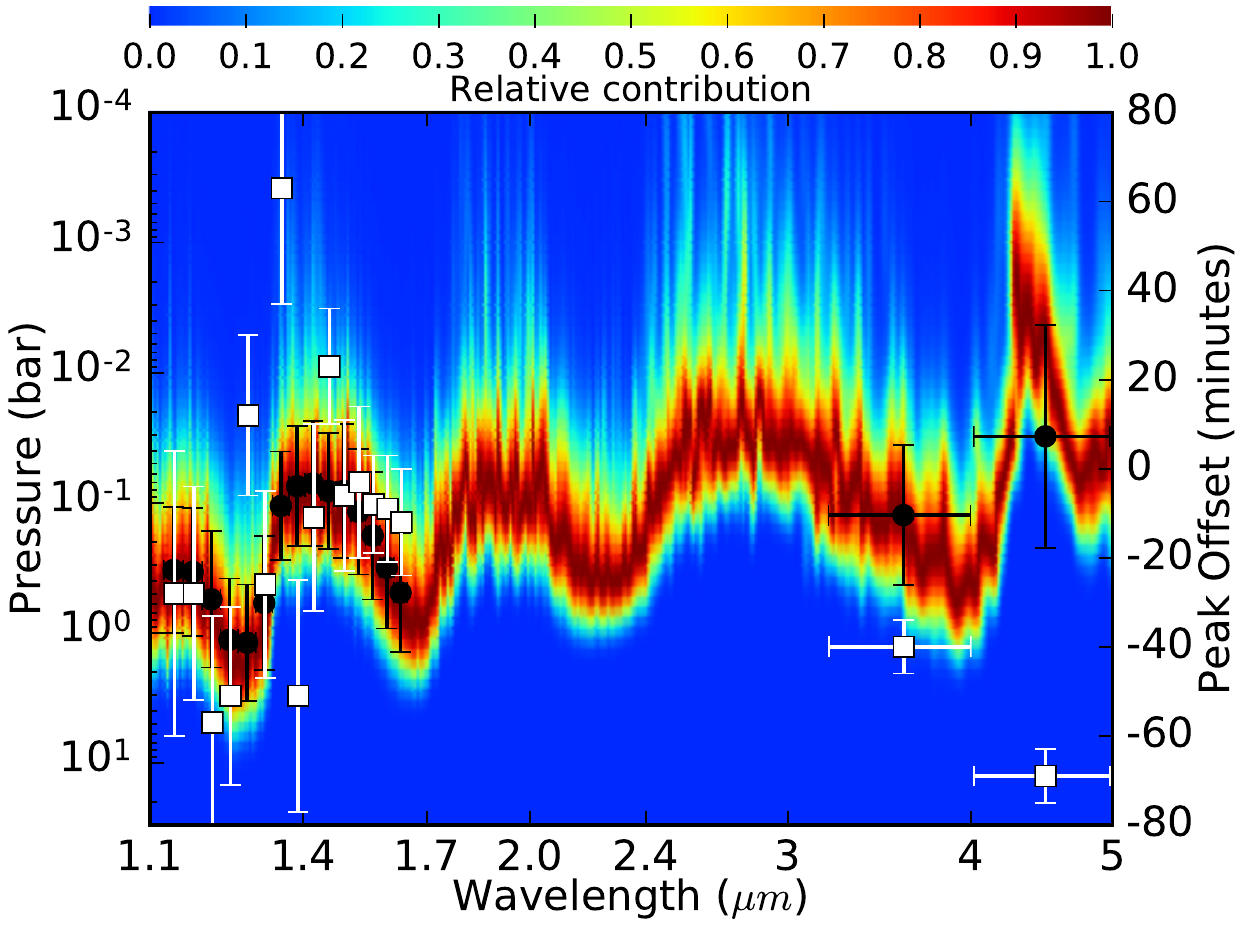}
\caption{\label{fig:contfunc}{
Dayside thermal emission contribution function of WASP-43b with phase curve peak offset.
Red indicates the pressure levels at which the optical depth is unity. These regions have the most significant
contribution to the wavelength-dependent emission. Blue indicates regions with negligible contribution to the total emission.  Black circles signify the mean pressure level at peak contribution in each spectrophotometric channel; white squares depict the phase-curve peak offsets from \citet{Stevenson2014c} and this work.  Horizontal and vertical error bars represent bandpass widths and 1$\sigma$ pressure level uncertainties, respectively.  The correlation between the dayside thermal emission contribution levels and phase-curve peak offsets seen in the WFC3 data does not extend to the IRAC data.
}}
\end{figure}

In \citet{Stevenson2014c}, we reported a trend between the dayside thermal emission contribution levels (the pressures at which our observations are sensitive to) and the measured phase-curve peak offsets.  Inside the water band (1.35 -- 1.6~{\microns}), WFC3 probed lower atmospheric pressures (relative to the other wavelengths) and we measured smaller phase-curve peak offsets.  Figure \ref{fig:contfunc} depicts WASP-43b's dayside contribution function based on our updated fit using all datasets.  The large offset at 4.5~{\microns} stands out as a clear outlier and, in general, the {\em Spitzer} points do not corroborate the reported trend.  This may be because {\em Spitzer}'s broad photometric channels can encompass several orders of magnitude in pressure, thus making them intractable to this type of measurement.  Alternatively, the {\em Spitzer} phase curve peak offsets may simply be unreliable.  Spectroscopic phase curve observations using {\em JWST}'s NIRCam or NIRSpec instruments will provide higher fidelity constraints at these wavelengths.

\subsection{3D General Circulation Models}
\label{sec:gcm}

We compare our {\em Spitzer} phase curves to those predicted by 3D GCMs \citep{Kataria2015}.  These are cloud-free models that, unlike the retrieval models in Section \ref{sec:atm}, have no additional parameter tuning once the initial conditions are set.  Using the SPARC/MITgcm \citep{Showman2009}, a state-of-the-art coupled radiation and circulation model, we explored the effects of composition, metallicity, and frictional drag (a crude parameterization of possible Lorentz forces) on the atmospheric circulation of WASP-43b.  For additional information, we refer the reader to \citet{Kataria2015}.

\begin{figure}[t]
\centering
\includegraphics[width=1.\linewidth]{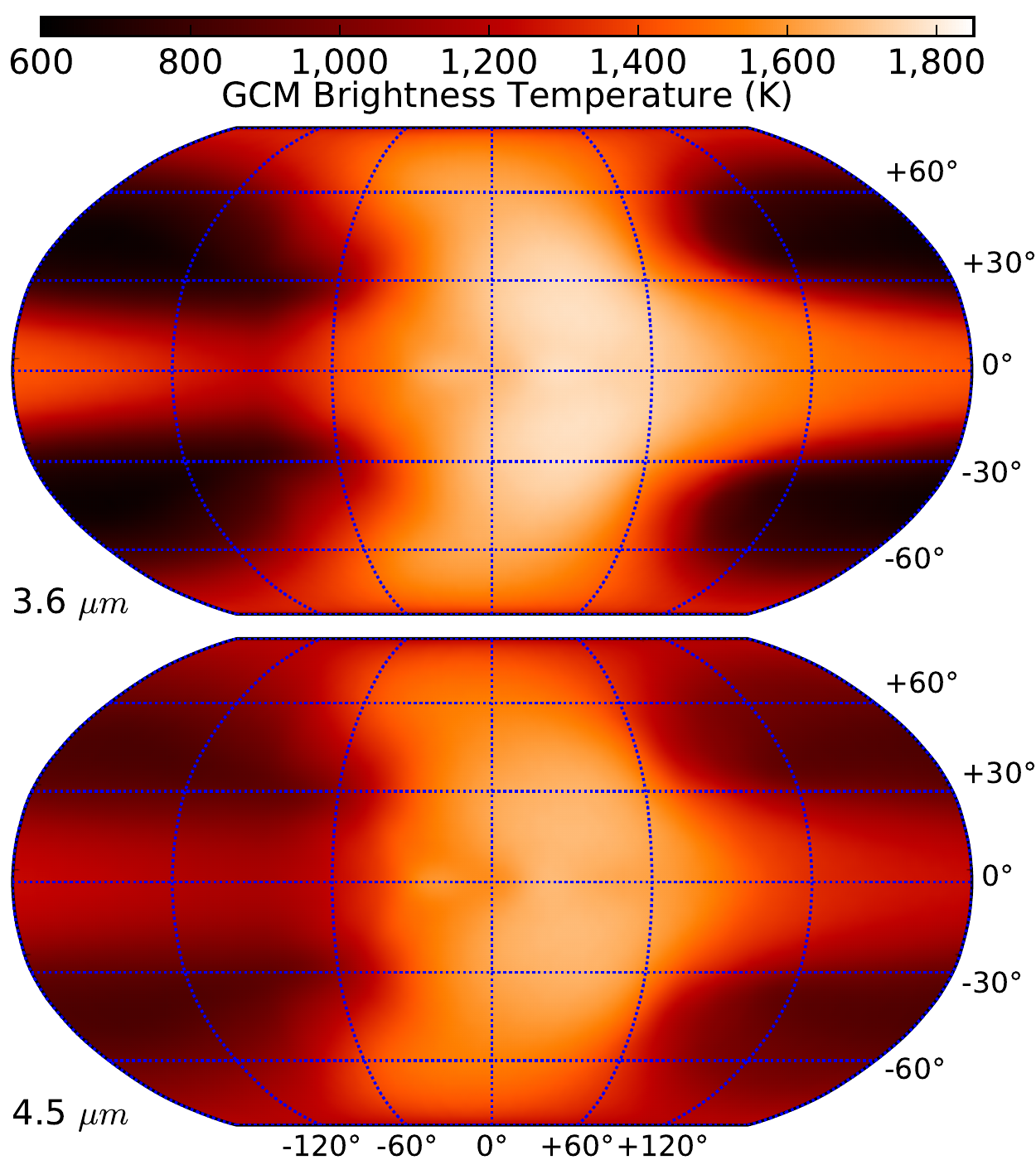}
\caption{\label{fig:gcmtb}{
GCM brightness temperature maps of WASP-43b at 3.6 and 4.5~{\microns} using solar metallicity models from \citet{Kataria2015}.  The pressure levels probed by both channels exhibit a hot and fast eastward equatorial jet, cold and slow westward mid-latitude flows, and moderate winds redistributing heat over the poles. 
}}
\end{figure}

Figure \ref{fig:gcmtb} depicts GCM brightness temperature maps of WASP-43b at 3.6 and 4.5~{\microns}.  The chevron shape of the heat distribution leads to a predicted eastward-shifted hotspot that is confirmed by the {\em Spitzer} phase curve observations.  In the models, heat is redistributed to the planet nightside via an equatorial superrotating jet and over both poles.  The mid-latitude regions at absolute longitudes $> 60$\degrees exhibit cooler temperatures and a slight westward flow.  This banded zonal flow is only seen at sufficiently high resolutions and in planets with relatively short orbital periods \citep[such as WASP-43b, ][]{Kataria2015, Kataria2016}.

\begin{figure}[t]
\centering
\includegraphics[width=1.0\linewidth]{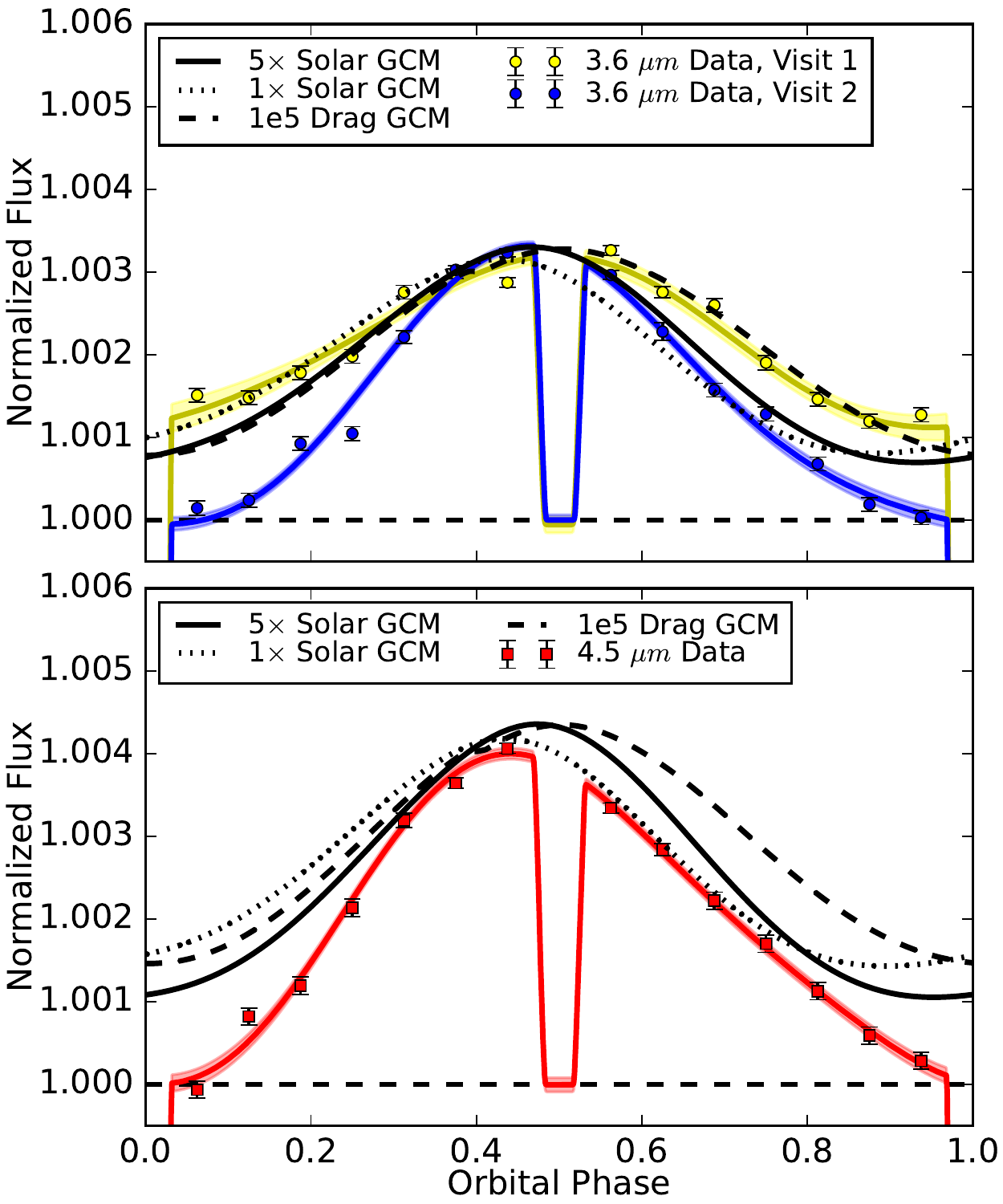}
\caption{\label{fig:gcm}{
GCM phase curves of WASP-43b at 3.6 (top) and 4.5 (bottom)~{\microns}.  
The black solid, dotted, and dashed lines depict predicted phase curves from cloud-free 3D GCMs of WASP-43b assuming $5\times$ solar metallicity, $1\times$ solar metallicity, and a frictional drag time constant of $10^5$~s, respectively \citep{Kataria2015}.  Colored symbols represent binned data that have been normalized with respect to the stellar flux.  The colored regions represent 1$\sigma$ uncertainty regions with respect to the median (colored curves).   The first visit at 3.6~{\microns} achieves good agreement with the frictional drag model at most orbital phases.  Data from the second 3.6~{\micron} visit best match the $5\times$ solar metallicity model on the planet dayside, but the model over-predicts the data on the nightside.  All models over-predict the 4.5~{\micron} nightside as well, which may suggest the presence of clouds on WASP-43b's nightside.
}}
\end{figure}

We calculate disk-integrated GCM phase curves following the procedures defined by \citet{Showman2008} and \citet{Fortney2006b} then, in Figure \ref{fig:gcm}, compare them to our measured phase curves from both 3.6~{\micron} visits and the 4.5~{\micron} visit.  Data from the first visit at 3.6~{\microns} achieve good agreement with the \ttt{5} frictional drag model at most orbital phases.  Conversely, the second (adopted) 3.6~{\micron} visit best matches the $5\times$ solar metallicity model on the planet dayside, but does not match any of the nightside models.  The GCMs also over-predict the nightside emission at 4.5~{\microns} and 1.1 -- 1.7~{\microns} (WFC3).  A plausible explanation for this discrepancy is the presence of clouds on WASP-43b's nightside that restrict our observations to higher altitudes where atmospheric temperatures are cooler \citep{Kataria2015}.  The first 3.6~{\micron} visit could then be explained by a temporary reduction in nightside cloud cover.  We caution, however, that this has not been proven and should not be taken as evidence for variability.  Rather, it is more likely that the first 3.6~{\micron} visit is simply yielding spurious results.

\subsection{Heat Redistribution Efficiency}

\citet{Perez-Becker2013} present evidence for a trend in which the hottest exoplanets exhibit inefficient heat redistribution (leading to strong day-night temperature contrasts) and cooler planets have increasingly more efficient heat transport (leading to more modest temperature contrasts).  Using an idealized, two-layer shallow water model to explain this trend, they theorize that the day-night temperature difference on synchronously rotating hot Jupiters is regulated by a wave-adjustment process.  Planetary-scale waves, which are triggered by the day-night heating contrast, propagate in longitude; vertical motions associated with these waves attempt to flatten isentropes and mute the day-night temperature difference.  When the radiative and frictional damping are weak, these waves can propagate from the dayside to the nightside, thus regulating the thermal structure and leading to modest day-night temperature differences.  However, when radiative and frictional damping are strong (manifesting as short radiative and/or frictional timescales), then the waves are damped before they can propagate across a hemisphere.  This suppresses the wave-adjustment mechanism and leads to large day-night temperature differences.  Because hotter planets generally should have shorter radiative time constants (and energy is deposited higher in their atmospheres), this mechanism explains the overall trend.

\citet{Komacek2016} extended this theory to the full 3D primitive equations.  Their model provides a more complete analytic prediction (in an idealized context) for the horizontal and vertical wind speeds and day-night temperature differences as a function of altitude for hot Jupiter atmospheres.  Nevertheless, the physical mechanism is essentially the same as that identified by \citet{Perez-Becker2013}.

\begin{figure}[t]
\centering
\includegraphics[width=1.\linewidth]{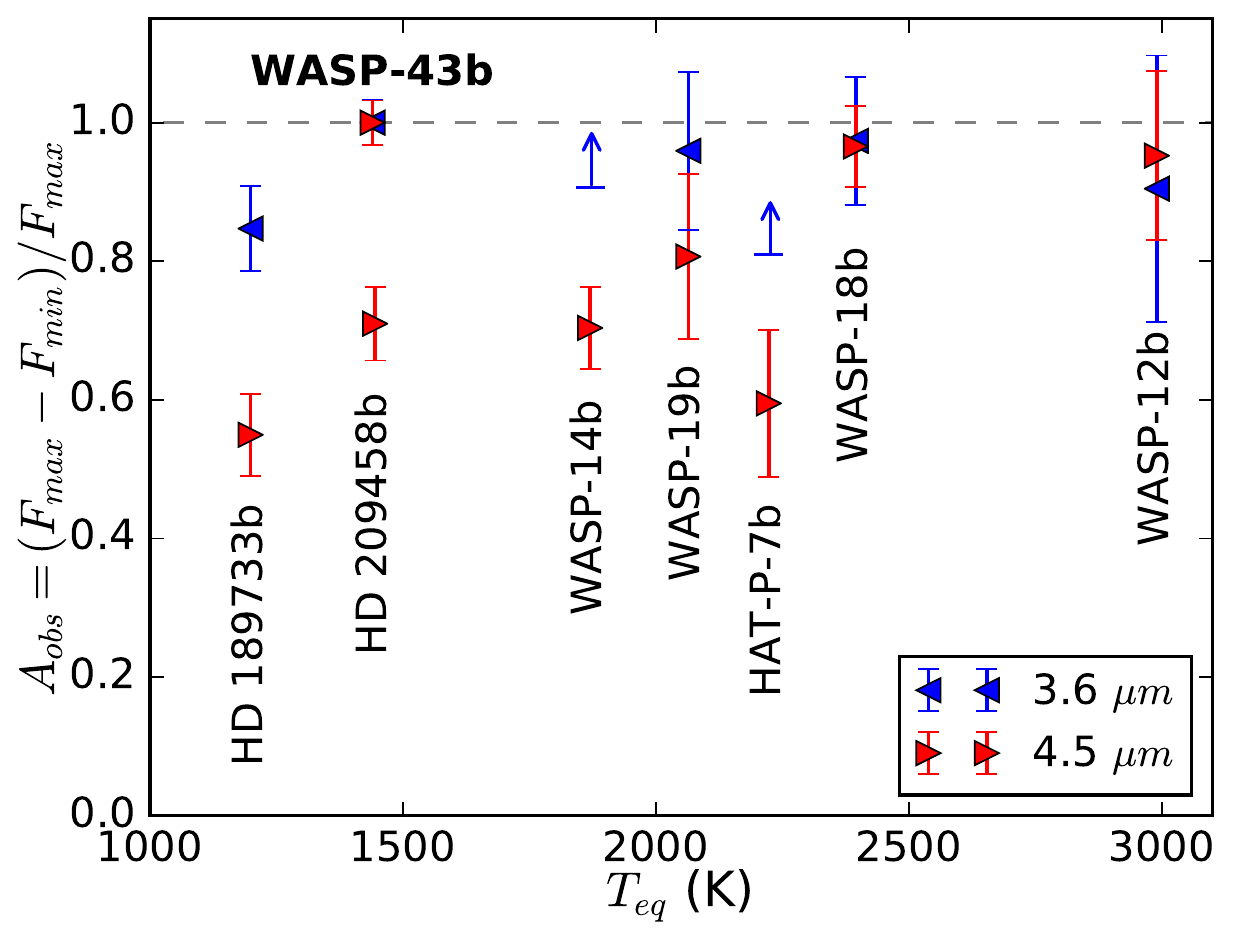}
\includegraphics[width=1.\linewidth]{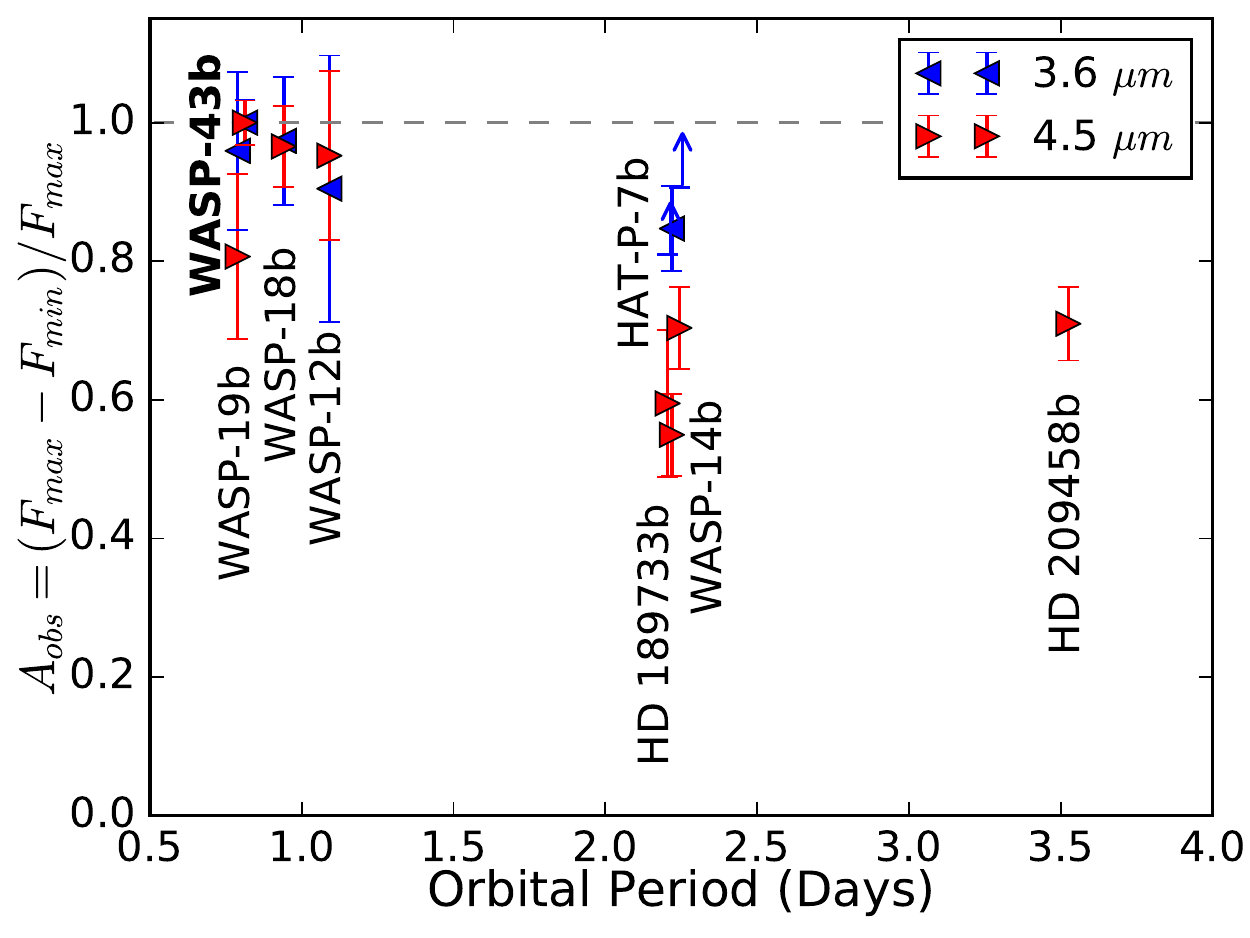}
\caption{\label{fig:heatRed}{
Heat redistribution efficiency, $A$\sb{obs}, versus equilibrium temperature (top) and orbital period (bottom) at 3.6 and 4.5~{\microns}.  Arrows depict $2\sigma$ lower limits; otherwise, the error bars represent $1\sigma$ uncertainties.  In the top panel, WASP-43b does not follow the trend reported by \citet{Perez-Becker2013}.  However, the current data support an apparent correlation between $A$\sb{obs} and planet orbital/rotational period (bottom panel).  We derive these results using data from \citet{Cowan2012, Knutson2012, Maxted2013, Zellem2014, Wong2015, Wong2016}.
}}
\end{figure}

This trend can be seen by plotting the observed fractional day-night flux difference, $A$\sb{obs}$=(F$\sb{max}$- F$\sb{min}$)/F$\sb{max}, as a function of equilibrium temperature, $T$\sb{eq}.  In the top panel of Figure \ref{fig:heatRed}, we plot $A$\sb{obs} versus $T$\sb{eq} for all of the exoplanets with published {\em Spitzer} phase curves at 3.6 and/or 4.5~{\microns}.  In general, we see an improvement in heat redistribution efficiency at lower temperatures in the 4.5~{\micron} bandpass.  However, WASP-43b does not follow this trend.

Despite not fitting the overall trend of $A$\sb{obs} versus $T$\sb{eq}, WASP-43b's large inferred day-night flux difference can still likely be understood within the context of the \citet{Perez-Becker2013} and \citet{Komacek2016} theories.  A fundamental result of these theories is that the criterion for the amplitude of the day-night temperature difference can be cast as a comparison between the day-night wave-propagation timescale and the radiative (and frictional) timescales.  Thus, if a thick cloud layer exists on the nightside of WASP-43b then it would shift the photosphere upward to a lower pressure where the radiative time constant is shorter.  In such a situation, the aforementioned theories predict a large day-night temperature difference, which is in agreement with our WASP-43b observations.  Essentially, the short-radiative-time constant models of \citet{Perez-Becker2013} may apply here, not because the irradiation is strong, but rather because obscuring clouds shift the photosphere to higher altitudes (lower pressures).  Under this scenario, the problem shifts toward understanding why WASP-43b exhibits a thick, high-altitude nightside cloud deck whereas other observed hot Jupiters (such as HD~189733b and HD~209458b) do not.

\subsection{A Dependence on Planet Rotation Rate}

To resolve this new problem, we first examine how the atmospheric flow depends on planet rotation rate.  For a synchronously rotating HD~209458b-like planet (which has a similar $T$\sb{eq} as WASP-43b), \citet{Showman2008} show that as the planet rotation rate increases (shorter orbital periods), the mid-latitude flow weakens and the superrotating equatorial jet gets narrower in latitude.  Next, we note that the wave-adjustment mechanism described by \citet{Perez-Becker2013} and \citet{Komacek2016} operates most efficiently within the equatorial wave guide.  Taken together, this means that for rapidly rotating planets like WASP-43b, the mid-to-high latitudes are farther outside the wave guide and can achieve cooler temperatures than the same latitudes on slowly rotating planets such as HD~209458b.  An example of this can be seen in Figure \ref{fig:gcmtb}, where the nightside mid-latitudes of the WASP-43b circulation models are significantly cooler than the equatorial region.

Therefore, we hypothesize that for more moderately-irradiated, faster-rotating planets such as WASP-43b, clouds/hazes are produced within the cooler, weaker mid-latitude flows and subsequently dispersed across the planet's nightside at high altitudes.  Atmospheric temperatures in these regions are favorable for the production of optically thick clouds/hazes, as shown by \citet{Morley2015}, and GCMs of planets with orbital periods of $\lesssim1$ day exhibit slow, slightly retrograde flows at mid-latitudes \citep{Kataria2015, Kataria2016} that could alter the relevant timescales.  Because of their longer orbital periods (slower rotation rates) and subsequently broader equatorial wave guides, planets such as HD~189733b and HD~209458b would not have the cooler and/or lower-wind-speed mid-latitude flows in which significant cloud formation could take place.  The inclusion of cloud physics in 3D GCMs should provide valuable insight into this theory.

In the lower panel of Figure \ref{fig:heatRed}, we plot $A$\sb{obs} versus orbital period, which is identical to the rotation rate for tidally locked planets, but not necessarily so for planets such as WASP-14b because of its slightly eccentric orbit.  The planet rotation rate is inversely proportional to the atmospheric Coriolis force, whose contribution should increase in importance with decreasing rotation rate.  Here we see that all of the planets with short ($\sim1$ day) orbital periods exhibit poor heat redistribution (high day-night contrasts) and all those with longer ($>2$ days) orbital periods exhibit more efficient heat redistribution (modest day-night contrasts) at 4.5~{\microns}.  Additional observations between these two regimes are needed to confirm this trend.

We can test the $A$\sb{obs} dependence on rotation rate by performing {\em Spitzer} phase curve observations of new exoplanet systems with specific temperatures and orbital periods.  In particular, observing multiple hot Jupiters with similar equilibrium temperatures and a range of orbital periods (between one and two days) should reveal a trend in the measured day-night contrast.  If we complement this work with additional sets of observations at other temperatures, we can more fully evaluate the connections between the measured day-night contrast, equilibrium temperature, and planet rotation rate, and ultimately better understand the day-night transport of energy in hot Jupiters.  Such a program would also provide a wealth of information about exoplanet atmospheres, including longitudinal constraints on atmospheric composition and thermal structure.

\section{SUMMARY}
\label{sec:concl}

Using the {\em Spitzer Space Telescope}, we obtained three broadband photometric phase curves of WASP-43b at 3.6 and 4.5~{\microns}.  We repeated the 3.6~{\micron} channel observation because the first visit exhibited strong nightside planetary emission that, when combined with the {\em HST}/WFC3 and 4.5~{\micron} nightside spectra, required invoking unphysical conditions in our atmospheric retrievals.  We initially suspected repointing inconsistencies between AORs (see Figure \ref{fig:pointingHist}, upper panel) as the source of the discrepant nightside flux, but tests using the second 3.6~{\micron} visit argued against this hypothesis.

Interestingly, the cloud-free GCM phase curves by \citet{Kataria2015} fit the data from the first 3.6~{\micron} visit at most orbital phases.  Therefore, the strong nightside emission could be explained by a temporary hole in a hypothesized nightside cloud deck.  Repeated phase curve observations would be necessary to search for additional signs of variability before any conclusions could be drawn.  Until then, the source of the nightside flux inconsistency remains a mystery.  For the remainder of this investigation, we adopt the phase curve data from the second 3.6~{\micron} visit.  

We detect strong asymmetries in the {\em Spitzer} phase curves that are similar in shape to the {\em HST}/WFC3 band-integrated phase curve.  This is the first multi-facility constraint on phase-dependent infrared emission for hot Jupiters.  The agreement between the results from the different instruments gives us confidence that these challenging observations can give reliable results with careful planning and analysis.  The measured eclipse depths are generally consistent with previous results \citep{Blecic2014}, but do show weak evidence for variability between epochs  ($<3\sigma$ confidence).  The near infrared slope in the measured transmission spectrum is consistent with a partially obscured signal due to the presence of clouds/hazes.  We find no evidence for orbital decay and determine a new, more accurate estimate of the planet's transit ephemeris, $T_0 = 2455528.86856(3)$ BJD\sb{TDB}, and orbital period, $p = 0.81347404(3)$ days.

Using the CHIMERA Bayesian retrieval suite on both the {\em HST} and {\em Spitzer} phase curve data, we perform independent atmospheric retrievals at 15 orbital phases.  The retrieved H\sb{2}O abundance shows some variation with orbital phase; however, it is unclear if this difference is physical or due to an unidentified bias within our models.  Using the more reliable dayside hemisphere (orbital phase = 0.28 $\to$ 0.72, excluding secondary eclipse) abundance of $\log$ [H\sb{2}O] $= -4.3{\pm}0.3$, we derive a precise metallicity constraint of 0.4 -- 1.7$\times$ solar at 1$\sigma$ confidence.  This value is consistent with our previous estimate \citep[0.4 -- 3.5$\times$,][]{Kreidberg2014b} based on the WFC3+IRAC transit and eclipse data alone.  The retrieved CO+CO\sb{2} abundance is constant with orbital phase.  We estimate a global log abundance of $-3.5{\pm}0.4$ and a nearly identical metallicity constraint of 0.3 -- 1.7$\times$ solar.  This is the first time that precise abundances have been determined from multiple molecular tracers for a transiting exoplanet. The consistency between the results is remarkable, and it bodes well for the {\em JWST} science ambitions of the exoplanet atmosphere community \citep{Beichman2014, Cowan2015, Stevenson2016c}.

The CH\sb{4} abundance varies depending on the number of thermal profiles in our model.  Using a single thermal profile, we retrieve an upper limit on the planet dayside and a bounded constraint near first and third quarters (see Figure \ref{fig:abundances}, panel (d)).  Using two thermal profiles to represent contributions from both a hot dayside and a cold nightside, the CH\sb{4} abundance at first quarter reverts to an upper limit that is consistent with the dayside constraint.  \citet{Feng2016} provide a more detailed discussion about the use of a second thermal profile and its impact at third quarter.

When we compare the {\em Spitzer} data to GCM phase curves, we achieve good fits on the planet dayside; however, the models over-predict emission on the planet nightside.  This discrepancy could be explained by the presence of optically thick clouds, which are not included in the GCMs.  However, one must then explain why other observed exoplanets with similar brightness temperatures more closely match their predicted nightside emission levels without the need for obscuring clouds \citep{Knutson2012, Zellem2014}.  

We hypothesize that exoplanet rotation rate may play an important, previously unknown role in the formation of a high-altitude nightside cloud deck.  As illustrated in the lower panel of Figure \ref{fig:heatRed}, relatively fast rotators ($\sim$1 day) have strong day-night contrasts (suggesting either poor heat redistribution or high-altitude nightside clouds) and slower rotators (>2 day) have more moderate day-night contrasts (favoring more efficient heat redistribution).  Additional phase curve observations targeting key exoplanets and the inclusion of cloud physics in 3D GCMs should provide valuable insight into this theory. 

WASP-43b's strong day-night contrast in all measured channels argues against a wavelength dependence in the heat redistribution efficiency and is inconsistent with the observational trend identified by \citet{Cowan2011} and \citet{Perez-Becker2013}.  Nevertheless, our result may still be consistent with theoretical predictions if WASP-43b has an unusually short radiative time constant, not because of strong irradiation, but rather due to a high-altitude cloud deck that shifts the nightside photosphere to low pressures \citep{Kataria2015}.  This would explain why WASP-43b's strong day-night contrast is consistent with other short-orbital period planets, which tend to be more highly irradiated.

\acknowledgments

We appreciate the thoughtful suggestions from the anonymous referee.  
We thank contributors to SciPy, Matplotlib, and the Python Programming Language, the free and open-source community, the NASA Astrophysics Data System, and the JPL Solar System Dynamics group for software and services.  
K.B.S. recognizes support from the Sagan Fellowship Program, supported by NASA and administered by the NASA Exoplanet Science Institute (NExScI).  J.L.B. acknowledges support from the David and Lucile Packard Foundation.
\\

\bibliography{ms}

\begin{thebibliography}{}
\expandafter\ifx\csname natexlab\endcsname\relax\def\natexlab#1{#1}\fi

\bibitem[{{Adams} {et~al.}(2010){Adams}, {L{\'o}pez-Morales}, {Elliot},
  {Seager}, \& {Osip}}]{Adams2010}
{Adams}, E.~R., {L{\'o}pez-Morales}, M., {Elliot}, J.~L., {Seager}, S., \&
  {Osip}, D.~J. 2010, \apj, 721, 1829

\bibitem[{{Agol} {et~al.}(2010){Agol}, {Cowan}, {Knutson}, {Deming}, {Steffen},
  {Henry}, \& {Charbonneau}}]{Agol2010}
{Agol}, E., {Cowan}, N.~B., {Knutson}, H.~A., {et~al.} 2010, \apj, 721, 1861

\bibitem[{{Ag{\'u}ndez} {et~al.}(2014){Ag{\'u}ndez}, {Parmentier}, {Venot},
  {Hersant}, \& {Selsis}}]{Agundez2014}
{Ag{\'u}ndez}, M., {Parmentier}, V., {Venot}, O., {Hersant}, F., \& {Selsis},
  F. 2014, \aap, 564, A73

\bibitem[{{Allard} {et~al.}(2000){Allard}, {Hauschildt}, \&
  {Schweitzer}}]{Allard2000}
{Allard}, F., {Hauschildt}, P.~H., \& {Schweitzer}, A. 2000, \apj, 539, 366

\bibitem[{{Beichman} {et~al.}(2014){Beichman}, {Benneke}, {Knutson}, {Smith},
  {Lagage}, {Dressing}, {Latham}, {Lunine}, {Birkmann}, {Ferruit}, {Giardino},
  {Kempton}, {Carey}, {Krick}, {Deroo}, {Mandell}, {Ressler}, {Shporer},
  {Swain}, {Vasisht}, {Ricker}, {Bouwman}, {Crossfield}, {Greene}, {Howell},
  {Christiansen}, {Ciardi}, {Clampin}, {Greenhouse}, {Sozzetti}, {Goudfrooij},
  {Hines}, {Keyes}, {Lee}, {McCullough}, {Robberto}, {Stansberry}, {Valenti},
  {Rieke}, {Rieke}, {Fortney}, {Bean}, {Kreidberg}, {Ehrenreich}, {Deming},
  {Albert}, {Doyon}, \& {Sing}}]{Beichman2014}
{Beichman}, C., {Benneke}, B., {Knutson}, H., {et~al.} 2014, \pasp, 126, 1134

\bibitem[{Bevington \& Robinson(2003)}]{Bevington2003}
Bevington, P.~R., \& Robinson, D.~K. 2003, {Data reduction and error analysis
  for the physical sciences; 3rd ed.} (New York, NY: McGraw-Hill)

\bibitem[{{Blecic} {et~al.}(2013){Blecic}, {Harrington}, {Madhusudhan},
  {Stevenson}, {Hardy}, {Blecic}, {Anderson}, {Hardin}, \&
  {Campo}}]{Blecic2013}
{Blecic}, J., {Harrington}, J., {Madhusudhan}, N., {et~al.} 2013, \apj, 779, 5

\bibitem[{{Blecic} {et~al.}(2014){Blecic}, {Harrington}, {Madhusudhan},
  {Stevenson}, {Hardy}, {Cubillos}, {Hardin}, {Bowman}, {Nymeyer}, {Anderson},
  {Hellier}, {Smith}, \& {Collier Cameron}}]{Blecic2014}
---. 2014, \apj, 781, 116

\bibitem[{{Burrows} {et~al.}(2008){Burrows}, {Ibgui}, \&
  {Hubeny}}]{Burrows2008}
{Burrows}, A., {Ibgui}, L., \& {Hubeny}, I. 2008, \apj, 682, 1277

\bibitem[{{Campo} {et~al.}(2011){Campo}, {Harrington}, {Hardy}, {Stevenson},
  {et~al.}}]{Campo2011}
{Campo}, C.~J., {Harrington}, J., {Hardy}, R.~A., {Stevenson}, K.~B., {et~al.}
  2011, \apj, 727, 125

\bibitem[{{Carter} \& {Winn}(2009)}]{Carter2009b}
{Carter}, J.~A., \& {Winn}, J.~N. 2009, \apj, 704, 51

\bibitem[{{Castelli} \& {Kurucz}(2004)}]{Kurucz2004}
{Castelli}, F., \& {Kurucz}, R.~L. 2004, ArXiv Astrophysics e-prints,
  arXiv:astro-ph/0405087

\bibitem[{{Charbonneau} {et~al.}(2005){Charbonneau}, {Allen}, {Megeath},
  {Torres}, {Alonso}, {Brown}, {Gilliland}, {Latham}, {Mandushev}, {O'Donovan},
  \& {Sozzetti}}]{Charbonneau2005}
{Charbonneau}, D., {Allen}, L.~E., {Megeath}, S.~T., {et~al.} 2005, \apj, 626,
  523

\bibitem[{{Chen} {et~al.}(2014){Chen}, {van Boekel}, {Wang}, {Nikolov},
  {Fortney}, {Seemann}, {Wang}, {Mancini}, \& {Henning}}]{Chen2014}
{Chen}, G., {van Boekel}, R., {Wang}, H., {et~al.} 2014, \aap, 563, A40

\bibitem[{{Claret}(2000)}]{Claret2000}
{Claret}, A. 2000, \aap, 363, 1081

\bibitem[{{Cooper} \& {Showman}(2006)}]{Cooper2006}
{Cooper}, C.~S., \& {Showman}, A.~P. 2006, \apj, 649, 1048

\bibitem[{{Cowan} \& {Agol}(2008)}]{Cowan2008}
{Cowan}, N.~B., \& {Agol}, E. 2008, \apjl, 678, L129

\bibitem[{{Cowan} \& {Agol}(2011)}]{Cowan2011}
---. 2011, \apj, 726, 82

\bibitem[{{Cowan} {et~al.}(2012){Cowan}, {Machalek}, {Croll}, {Shekhtman},
  {Burrows}, {Deming}, {Greene}, \& {Hora}}]{Cowan2012}
{Cowan}, N.~B., {Machalek}, P., {Croll}, B., {et~al.} 2012, \apj, 747, 82

\bibitem[{{Cowan} {et~al.}(2015){Cowan}, {Greene}, {Angerhausen}, {Batalha},
  {Clampin}, {Col{\'o}n}, {Crossfield}, {Fortney}, {Gaudi}, {Harrington},
  {Iro}, {Lillie}, {Linsky}, {Lopez-Morales}, {Mandell}, {Stevenson}, \&
  {ExoPAG SAG-10}}]{Cowan2015}
{Cowan}, N.~B., {Greene}, T., {Angerhausen}, D., {et~al.} 2015, \pasp, 127, 311

\bibitem[{{Crossfield}(2015)}]{Crossfield2015b}
{Crossfield}, I.~J.~M. 2015, \pasp, 127, 941

\bibitem[{{Cubillos} {et~al.}(2013){Cubillos}, {Harrington}, {Madhusudhan},
  {Stevenson}, {Hardy}, {Blecic}, {Anderson}, {Hardin}, \&
  {Campo}}]{Cubillos2013}
{Cubillos}, P., {Harrington}, J., {Madhusudhan}, N., {et~al.} 2013, \apj, 768,
  42

\bibitem[{{Deming} {et~al.}(2015){Deming}, {Knutson}, {Kammer}, {Fulton},
  {Ingalls}, {Carey}, {Burrows}, {Fortney}, {Todorov}, {Agol}, {Cowan},
  {Desert}, {Fraine}, {Langton}, {Morley}, \& {Showman}}]{Deming2015}
{Deming}, D., {Knutson}, H., {Kammer}, J., {et~al.} 2015, \apj, 805, 132

\bibitem[{{Diamond-Lowe} {et~al.}(2014){Diamond-Lowe}, {Stevenson}, {Bean},
  {Line}, \& {Fortney}}]{HDL2014}
{Diamond-Lowe}, H., {Stevenson}, K.~B., {Bean}, J.~L., {Line}, M.~R., \&
  {Fortney}, J.~J. 2014, \apj, 796, 66

\bibitem[{{Fazio} {et~al.}(2004){Fazio}, {Hora}, {Allen}, {et~al.}}]{IRAC}
{Fazio}, G.~G., {Hora}, J.~L., {Allen}, L.~E., {et~al.} 2004, Astrophy. J.
  Suppl. Ser., 154, 10

\bibitem[{{Feng} {et~al.}(2016){Feng}, {Line}, {Fortney}, {Stevenson}, {Bean},
  {Kreidberg}, \& {Parmentier}}]{Feng2016}
{Feng}, Y.~K., {Line}, M.~R., {Fortney}, J.~J., {et~al.} 2016, \apj, 829, 52

\bibitem[{{Fortney} {et~al.}(2006){Fortney}, {Cooper}, {Showman}, {Marley}, \&
  {Freedman}}]{Fortney2006b}
{Fortney}, J.~J., {Cooper}, C.~S., {Showman}, A.~P., {Marley}, M.~S., \&
  {Freedman}, R.~S. 2006, \apj, 652, 746

\bibitem[{{Fortney} {et~al.}(2008){Fortney}, {Lodders}, {Marley}, \&
  {Freedman}}]{Fortney2008}
{Fortney}, J.~J., {Lodders}, K., {Marley}, M.~S., \& {Freedman}, R.~S. 2008,
  \apj, 678, 1419

\bibitem[{{Gillon} {et~al.}(2012){Gillon}, {Triaud}, {Fortney}, {Demory},
  {Jehin}, {Lendl}, {Magain}, {Kabath}, {Queloz}, {Alonso}, {Anderson},
  {Collier Cameron}, {Fumel}, {Hebb}, {Hellier}, {Lanotte}, {Maxted},
  {Mowlavi}, \& {Smalley}}]{Gillon2012}
{Gillon}, M., {Triaud}, A.~H.~M.~J., {Fortney}, J.~J., {et~al.} 2012, \aap,
  542, A4

\bibitem[{{Gordon} \& {McBride}(1994)}]{Gordon1994}
{Gordon}, S., \& {McBride}, B.~J. 1994, NASA Reference Publ. 1311

\bibitem[{{Hansen} {et~al.}(2014){Hansen}, {Schwartz}, \& {Cowan}}]{Hansen2014}
{Hansen}, C.~J., {Schwartz}, J.~C., \& {Cowan}, N.~B. 2014, \mnras, 444, 3632

\bibitem[{{Harrington} {et~al.}(2007){Harrington}, {Luszcz}, {Seager},
  {Deming}, \& {Richardson}}]{Harrington2007}
{Harrington}, J., {Luszcz}, S., {Seager}, S., {Deming}, D., \& {Richardson},
  L.~J. 2007, Nature, 447, 691

\bibitem[{{Hellier} {et~al.}(2011){Hellier}, {Anderson}, {Collier Cameron},
  {Gillon}, {Jehin}, {Lendl}, {Maxted}, {Pepe}, {Pollacco}, {Queloz},
  {S{\'e}gransan}, {Smalley}, {Smith}, {Southworth}, {Triaud}, {Udry}, \&
  {West}}]{Hellier2011}
{Hellier}, C., {Anderson}, D.~R., {Collier Cameron}, A., {et~al.} 2011, \aap,
  535, L7

\bibitem[{{Hoyer} {et~al.}(2016){Hoyer}, {Pall{\'e}}, {Dragomir}, \&
  {Murgas}}]{Hoyer2016}
{Hoyer}, S., {Pall{\'e}}, E., {Dragomir}, D., \& {Murgas}, F. 2016, \aj, 151,
  137

\bibitem[{{Ingalls} {et~al.}(2012){Ingalls}, {Krick}, {Carey}, {Laine},
  {Surace}, {Glaccum}, {Grillmair}, \& {Lowrance}}]{Ingalls2012}
{Ingalls}, J.~G., {Krick}, J.~E., {Carey}, S.~J., {et~al.} 2012, in Society of
  Photo-Optical Instrumentation Engineers (SPIE) Conference Series, Vol. 8442,
  Society of Photo-Optical Instrumentation Engineers (SPIE) Conference Series,
  1

\bibitem[{{Ingalls} {et~al.}(2016){Ingalls}, {Krick}, {Carey}, {Stauffer},
  {Lowrance}, {Grillmair}, {Buzasi}, {Deming}, {Diamond-Lowe}, {Evans},
  {Morello}, {Stevenson}, {Wong}, {Capak}, {Glaccum}, {Laine}, {Surace}, \&
  {Storrie-Lombardi}}]{Ingalls2016}
{Ingalls}, J.~G., {Krick}, J.~E., {Carey}, S.~J., {et~al.} 2016, \aj, 152, 44

\bibitem[{{Jiang} {et~al.}(2016){Jiang}, {Lai}, {Savushkin}, {Mkrtichian},
  {Antonyuk}, {Griv}, {Hsieh}, \& {Yeh}}]{Jiang2016}
{Jiang}, I.-G., {Lai}, C.-Y., {Savushkin}, A., {et~al.} 2016, \aj, 151, 17

\bibitem[{{Kataria} {et~al.}(2015){Kataria}, {Showman}, {Fortney}, {Stevenson},
  {Line}, {Kreidberg}, {Bean}, \& {D{\'e}sert}}]{Kataria2015}
{Kataria}, T., {Showman}, A.~P., {Fortney}, J.~J., {et~al.} 2015, \apj, 801, 86

\bibitem[{{Kataria} {et~al.}(2016){Kataria}, {Sing}, {Lewis}, {Visscher},
  {Showman}, {Fortney}, \& {Marley}}]{Kataria2016}
{Kataria}, T., {Sing}, D.~K., {Lewis}, N.~K., {et~al.} 2016, \apj, 821, 9

\bibitem[{{Knutson} {et~al.}(2011){Knutson}, {Madhusudhan}, {Cowan},
  {Christiansen}, {Agol}, {Deming}, {D{\'e}sert}, {Charbonneau}, {Henry},
  {Homeier}, {Langton}, {Laughlin}, \& {Seager}}]{Knutson2011}
{Knutson}, H.~A., {Madhusudhan}, N., {Cowan}, N.~B., {et~al.} 2011, \apj, 735,
  27

\bibitem[{{Knutson} {et~al.}(2012){Knutson}, {Lewis}, {Fortney}, {Burrows},
  {Showman}, {Cowan}, {Agol}, {Aigrain}, {Charbonneau}, {Deming}, {D{\'e}sert},
  {Henry}, {Langton}, \& {Laughlin}}]{Knutson2012}
{Knutson}, H.~A., {Lewis}, N., {Fortney}, J.~J., {et~al.} 2012, \apj, 754, 22

\bibitem[{{Komacek} \& {Showman}(2016)}]{Komacek2016}
{Komacek}, T.~D., \& {Showman}, A.~P. 2016, \apj, 821, 16

\bibitem[{{Kreidberg} {et~al.}(2014){Kreidberg}, {Bean}, {D{\'e}sert}, {Line},
  {Fortney}, {Madhusudhan}, {Stevenson}, {Showman}, {Charbonneau},
  {McCullough}, {Seager}, {Burrows}, {Henry}, {Williamson}, {Kataria}, \&
  {Homeier}}]{Kreidberg2014b}
{Kreidberg}, L., {Bean}, J.~L., {D{\'e}sert}, J.-M., {et~al.} 2014, \apjl, 793,
  L27

\bibitem[{{Lewis} {et~al.}(2013){Lewis}, {Knutson}, {Showman}, {Cowan},
  {Laughlin}, {Burrows}, {Deming}, {Crepp}, {Mighell}, {Agol}, {Bakos},
  {Charbonneau}, {D{\'e}sert}, {Fischer}, {Fortney}, {Hartman}, {Hinkley},
  {Howard}, {Johnson}, {Kao}, {Langton}, \& {Marcy}}]{Lewis2013}
{Lewis}, N.~K., {Knutson}, H.~A., {Showman}, A.~P., {et~al.} 2013, \apj, 766,
  95

\bibitem[{{Liddle}(2007)}]{Liddle2008}
{Liddle}, A.~R. 2007, Mon. Not. R. Astron. Soc., 377, L74

\bibitem[{{Line} {et~al.}(2014){Line}, {Knutson}, {Wolf}, \&
  {Yung}}]{Line2014-C/O}
{Line}, M.~R., {Knutson}, H., {Wolf}, A.~S., \& {Yung}, Y.~L. 2014, \apj, 783,
  70

\bibitem[{{Line} \& {Parmentier}(2016)}]{Line2016a}
{Line}, M.~R., \& {Parmentier}, V. 2016, \apj, 820, 78

\bibitem[{{Line} {et~al.}(2013){Line}, {Wolf}, {Zhang}, {Knutson}, {Kammer},
  {Ellison}, {Deroo}, {Crisp}, \& {Yung}}]{Line2013a}
{Line}, M.~R., {Wolf}, A.~S., {Zhang}, X., {et~al.} 2013, \apj, 775, 137

\bibitem[{{Line} {et~al.}(2016){Line}, {Stevenson}, {Bean}, {Desert},
  {Fortney}, {Kreidberg}, {Madhusudhan}, {Showman}, \&
  {Diamond-Lowe}}]{Line2016b}
{Line}, M.~R., {Stevenson}, K.~B., {Bean}, J., {et~al.} 2016, \aj, 152, 203

\bibitem[{{Lodders}(2002)}]{Lodders2002}
{Lodders}, K. 2002, \apj, 577, 974

\bibitem[{{Lust} {et~al.}(2014){Lust}, {Britt}, {Harrington}, {Nymeyer},
  {Stevenson}, {Ross}, {Bowman}, \& {Fraine}}]{Lust2014}
{Lust}, N.~B., {Britt}, D., {Harrington}, J., {et~al.} 2014, \pasp, 126, 1092

\bibitem[{{Mandel} \& {Agol}(2002)}]{Mandel2002}
{Mandel}, K., \& {Agol}, E. 2002, \apj, 580, L171

\bibitem[{{Maxted} {et~al.}(2013){Maxted}, {Anderson}, {Doyle}, {Gillon},
  {Harrington}, {Iro}, {Jehin}, {Lafreni{\`e}re}, {Smalley}, \&
  {Southworth}}]{Maxted2013}
{Maxted}, P.~F.~L., {Anderson}, D.~R., {Doyle}, A.~P., {et~al.} 2013, \mnras,
  428, 2645

\bibitem[{{Morley} {et~al.}(2015){Morley}, {Fortney}, {Marley}, {Zahnle},
  {Line}, {Kempton}, {Lewis}, \& {Cahoy}}]{Morley2015}
{Morley}, C.~V., {Fortney}, J.~J., {Marley}, M.~S., {et~al.} 2015, \apj, 815,
  110

\bibitem[{{Murgas} {et~al.}(2014){Murgas}, {Pall{\'e}}, {Zapatero Osorio},
  {Nortmann}, {Hoyer}, \& {Cabrera-Lavers}}]{Murgas2014}
{Murgas}, F., {Pall{\'e}}, E., {Zapatero Osorio}, M.~R., {et~al.} 2014, \aap,
  563, A41

\bibitem[{{Parmentier} {et~al.}(2016){Parmentier}, {Fortney}, {Showman},
  {Morley}, \& {Marley}}]{Parmentier2016}
{Parmentier}, V., {Fortney}, J.~J., {Showman}, A.~P., {Morley}, C., \&
  {Marley}, M.~S. 2016, \apj, 828, 22

\bibitem[{{Perez-Becker} \& {Showman}(2013)}]{Perez-Becker2013}
{Perez-Becker}, D., \& {Showman}, A.~P. 2013, \apj, 776, 134

\bibitem[{{Perna} {et~al.}(2012){Perna}, {Heng}, \& {Pont}}]{Perna2012}
{Perna}, R., {Heng}, K., \& {Pont}, F. 2012, \apj, 751, 59

\bibitem[{{Pont} {et~al.}(2013){Pont}, {Sing}, {Gibson}, {Aigrain}, {Henry}, \&
  {Husnoo}}]{Pont2013}
{Pont}, F., {Sing}, D.~K., {Gibson}, N.~P., {et~al.} 2013, \mnras, 432, 2917

\bibitem[{{Ricci} {et~al.}(2015){Ricci}, {Ram{\'o}n-Fox}, {Ayala-Loera},
  {Michel}, {Navarro-Meza}, {Fox-Machado}, {Reyes-Ruiz}, {Sevilla}, \&
  {Curiel}}]{Ricci2015}
{Ricci}, D., {Ram{\'o}n-Fox}, F.~G., {Ayala-Loera}, C., {et~al.} 2015, \pasp,
  127, 143

\bibitem[{{Rogers} {et~al.}(2013){Rogers}, {L{\'o}pez-Morales}, {Apai}, \&
  {Adams}}]{Rogers2013}
{Rogers}, J., {L{\'o}pez-Morales}, M., {Apai}, D., \& {Adams}, E. 2013, \apj,
  767, 64

\bibitem[{{Showman} {et~al.}(2008){Showman}, {Cooper}, {Fortney}, \&
  {Marley}}]{Showman2008}
{Showman}, A.~P., {Cooper}, C.~S., {Fortney}, J.~J., \& {Marley}, M.~S. 2008,
  \apj, 682, 559

\bibitem[{{Showman} {et~al.}(2009){Showman}, {Fortney}, {Lian}, {Marley},
  {Freedman}, {Knutson}, \& {Charbonneau}}]{Showman2009}
{Showman}, A.~P., {Fortney}, J.~J., {Lian}, Y., {et~al.} 2009, \apj, 699, 564

\bibitem[{{Sing} {et~al.}(2016){Sing}, {Fortney}, {Nikolov}, {Wakeford},
  {Kataria}, {Evans}, {Aigrain}, {Ballester}, {Burrows}, {Deming},
  {D{\'e}sert}, {Gibson}, {Henry}, {Huitson}, {Knutson}, {Etangs}, {Pont},
  {Showman}, {Vidal-Madjar}, {Williamson}, \& {Wilson}}]{Sing2016}
{Sing}, D.~K., {Fortney}, J.~J., {Nikolov}, N., {et~al.} 2016, \nat, 529, 59

\bibitem[{{Stevenson}(2016)}]{Stevenson2016b}
{Stevenson}, K.~B. 2016, \apjl, 817, L16

\bibitem[{{Stevenson} {et~al.}(2012){Stevenson}, {Harrington}, {Fortney},
  {Loredo}, {et~al.}}]{Stevenson2011}
{Stevenson}, K.~B., {Harrington}, J., {Fortney}, J.~J., {Loredo}, T.~J.,
  {et~al.} 2012, \apj, 754, 136

\bibitem[{{Stevenson} {et~al.}(2010){Stevenson}, {Harrington}, {Nymeyer},
  {Madhusudhan}, {et~al.}}]{Stevenson2010}
{Stevenson}, K.~B., {Harrington}, J., {Nymeyer}, S., {Madhusudhan}, N.,
  {et~al.} 2010, \nat, 464, 1161

\bibitem[{{Stevenson} {et~al.}(2014){Stevenson}, {Desert}, {Line}, {Bean},
  {Fortney}, {Showman}, {Kataria}, {Kreidberg}, {McCullough}, {Henry},
  {Charbonneau}, {Burrows}, {Seager}, {Madhusudhan}, {Williamson}, \&
  {Homeier}}]{Stevenson2014c}
{Stevenson}, K.~B., {Desert}, J.-M., {Line}, M.~R., {et~al.} 2014, Science,
  346, 838

\bibitem[{{Stevenson} {et~al.}(2016){Stevenson}, {Lewis}, {Bean}, {Beichman},
  {Fraine}, {Kilpatrick}, {Krick}, {Lothringer}, {Mandell}, {Valenti}, {Agol},
  {Angerhausen}, {Barstow}, {Birkmann}, {Burrows}, {Cowan}, {Crouzet},
  {Cubillos}, {Curry}, {Dalba}, {de Wit}, {Deming}, {Desert}, {Doyon},
  {Dragomir}, {Ehrenreich}, {Fortney}, {Garcia Munoz}, {Gibson}, {Gizis},
  {Greene}, {Harrington}, {Heng}, {Kataria}, {Kempton}, {Knutson}, {Kreidberg},
  {Lafreniere}, {Lagage}, {Line}, {Lopez-Morales}, {Madhusudhan}, {Morley},
  {Rocchetto}, {Schlawin}, {Shkolnik}, {Shporer}, {Sing}, {Todorov}, {Tucker},
  \& {Wakeford}}]{Stevenson2016c}
{Stevenson}, K.~B., {Lewis}, N.~K., {Bean}, J.~L., {et~al.} 2016, \pasp, 128,
  094401

\bibitem[{{Sudarsky} {et~al.}(2003){Sudarsky}, {Burrows}, \&
  {Hubeny}}]{Sudarsky2003}
{Sudarsky}, D., {Burrows}, A., \& {Hubeny}, I. 2003, \apj, 588, 1121

\bibitem[{ter Braak \& Vrugt(2008)}]{terBraak2008}
ter Braak, C., \& Vrugt, J. 2008, Statistics and Computing, 18, 435

\bibitem[{{Wang} {et~al.}(2013){Wang}, {van Boekel}, {Madhusudhan}, {Chen},
  {Zhao}, \& {Henning}}]{Wang2013}
{Wang}, W., {van Boekel}, R., {Madhusudhan}, N., {et~al.} 2013, \apj, 770, 70

\bibitem[{{Wong} {et~al.}(2015){Wong}, {Knutson}, {Lewis}, {Kataria},
  {Burrows}, {Fortney}, {Schwartz}, {Agol}, {Cowan}, {Deming}, {D{\'e}sert},
  {Fulton}, {Howard}, {Langton}, {Laughlin}, {Showman}, \&
  {Todorov}}]{Wong2015}
{Wong}, I., {Knutson}, H.~A., {Lewis}, N.~K., {et~al.} 2015, \apj, 811, 122

\bibitem[{{Wong} {et~al.}(2016){Wong}, {Knutson}, {Kataria}, {Lewis},
  {Burrows}, {Fortney}, {Schwartz}, {Shporer}, {Agol}, {Cowan}, {Deming},
  {D{\'e}sert}, {Fulton}, {Howard}, {Langton}, {Laughlin}, {Showman}, \&
  {Todorov}}]{Wong2016}
{Wong}, I., {Knutson}, H.~A., {Kataria}, T., {et~al.} 2016, \apj, 823, 122

\bibitem[{{Zellem} {et~al.}(2014){Zellem}, {Lewis}, {Knutson}, {Griffith},
  {Showman}, {Fortney}, {Cowan}, {Agol}, {Burrows}, {Charbonneau}, {Deming},
  {Laughlin}, \& {Langton}}]{Zellem2014}
{Zellem}, R.~T., {Lewis}, N.~K., {Knutson}, H.~A., {et~al.} 2014, \apj, 790, 53

\end{thebibliography}

\clearpage

\begin{sidewaystable}
\centering
\caption{\label{tab:spectra} 
Planet Emission With Phase-Independent Uncertainties}
\begin{tabular}{cccccccccccccccccc}
    \hline
    \hline
    Orbital & 1.1425 & 1.1775 & 1.2125 & 1.2475 & 1.2825 & 1.3175 & 1.3525 & 1.3875 & 1.4225 & 1.4575 & 1.4925 & 1.4275 & 1.5625 & 1.5975 & 1.6325 & 3.6 & 4.5 \\
    Phase   &({\microns})&({\microns})&({\microns})&({\microns})&({\microns})&({\microns})&({\microns})&({\microns})&({\microns})
            &({\microns})&({\microns})&({\microns})&({\microns})&({\microns})&({\microns})&({\microns})&({\microns})\\
    \hline
    0.0625 &  60(66) &  55(61) &  66(58) &  86(56) &  53(57) &  90(53) &   2(55) &  29(52) &  35(56) &  -3(56) &  24(56) &  -3(55) &  22(58) &  48(58) &  87(63) & -13(103) &  95(133) \\ 
    0.1250 & 103(67) & 105(61) & 125(59) & 154(56) &  92(57) & 144(53) &  19(55) &  74(52) &  71(56) &  33(56) &  74(56) &  50(55) &  91(58) & 116(58) & 167(63) & 235(105) & 524(133) \\ 
    0.1875 & 161(71) & 176(65) & 201(62) & 242(60) & 158(61) & 219(57) &  60(58) & 132(55) & 124(60) &  95(59) & 149(59) & 129(58) & 195(62) & 218(61) & 283(67) & 735(103) & 1302(136) \\ 
    0.2500 & 224(63) & 253(58) & 278(55) & 330(53) & 235(54) & 298(50) & 117(52) & 192(48) & 181(53) & 169(53) & 233(52) & 221(52) & 309(55) & 331(54) & 413(60) & 1458(103) & 2242(134) \\ 
    0.3125 & 283(69) & 326(64) & 347(61) & 407(59) & 314(59) & 373(56) & 183(57) & 247(54) & 236(59) & 246(58) & 315(58) & 311(57) & 422(61) & 443(60) & 539(66) & 2245(100) & 3145(119) \\ 
    0.3750 & 329(66) & 383(61) & 396(58) & 464(56) & 384(57) & 433(53) & 249(55) & 288(51) & 281(56) & 315(55) & 383(55) & 388(55) & 515(58) & 536(57) & 644(63) & 2909(79) & 3768(103) \\ 
    0.4375 & 355(59) & 417(53) & 419(52) & 491(49) & 433(50) & 467(47) & 303(48) & 309(44) & 309(49) & 364(49) & 426(48) & 439(48) & 575(51) & 595(51) & 708(56) & 3281(77) & 4000(103) \\ 
    0.5000 & 367(45) & 431(39) & 414(38) & 482(36) & 460(37) & 473(33) & 353(34) & 313(30) & 320(36) & 394(36) & 439(33) & 458(35) & 595(36) & 614(37) & 732(42) & 3231(60) & 3827(84) \\ 
    0.5625 & 335(61) & 399(55) & 375(53) & 441(51) & 445(52) & 444(48) & 349(50) & 280(46) & 298(51) & 379(51) & 420(50) & 437(50) & 563(52) & 583(52) & 691(58) & 2881(80) & 3389(103) \\ 
    0.6250 & 293(65) & 349(59) & 316(57) & 373(55) & 405(55) & 391(52) & 335(53) & 236(50) & 262(55) & 343(54) & 370(54) & 384(53) & 493(56) & 514(56) & 610(61) & 2285(121) & 2799(109) \\ 
    0.6875 & 237(71) & 281(65) & 242(63) & 289(61) & 342(61) & 317(57) & 295(59) & 179(56) & 211(60) & 283(59) & 298(60) & 307(59) & 393(62) & 415(62) & 497(67) & 1625(103) & 2204(133) \\ 
    0.7500 & 174(66) & 204(61) & 164(58) & 200(56) & 266(57) & 238(53) & 237(55) & 119(51) & 154(56) & 209(55) & 214(55) & 215(55) & 278(58) & 301(57) & 368(63) & 1054(103) & 1640(134) \\ 
    0.8125 & 115(68) & 131(63) &  96(60) & 121(58) & 186(58) & 162(55) & 171(56) &  64(53) &  99(57) & 132(57) & 131(57) & 124(56) & 165(59) & 189(59) & 242(64) & 617(103) & 1126(134) \\ 
    0.8750 &  67(63) &  71(57) &  45(55) &  63(52) & 114(53) & 100(50) & 103(51) &  22(48) &  52(53) &  61(52) &  61(52) &  45(51) &  68(54) &  94(54) & 135(59) & 299(103) & 645(133) \\ 
    0.9375 &  41(70) &  36(64) &  22(62) &  37(59) &  66(60) &  67(56) &  47(58) &   1(55) &  25(59) &  12(58) &  17(59) &  -5(58) &   9(61) &  36(61) &  70(66) &  83(103) & 247(133) \\
    \hline
\end{tabular}
\tablenotetext{0}{Fluxes are in units of ppm.  Values within the parentheses represent uncertainties.  We report secondary eclipse depths and uncertainties at 0.5 orbital phase.  The 3.6~{\micron} values originate from the second visit.}
\end{sidewaystable}

\end{document}